\documentclass[authoryear,12pt]{elsarticle}
\makeatletter
\def\ps@pprintTitle{%
 \let\@oddhead\@empty
 \let\@evenhead\@empty
 \def\@oddfoot{\centerline{\thepage}}%
 \let\@evenfoot\@oddfoot}
\makeatother

\usepackage{amssymb}

\usepackage{graphicx}
\usepackage{epstopdf}
\usepackage{bbm}
\usepackage{amsmath}
\usepackage{amsthm}
\usepackage[margin=1in]{geometry} 
\usepackage{url}

\usepackage{bm}

\newcommand{\mc}[1]{\ensuremath{\mathcal{#1}}}

\newtheorem{property}{Property}

\newtheorem{theorem}{Theorem}

\newtheorem{definition}{Definition}
\newtheorem{proposition}{Proposition}
\newtheorem{corollary}{Corollary}

\begin{document}
\begin{frontmatter} 
\title{Strategic Investment in Protection in Networked Systems\tnoteref{t1} \\ \ \\
\normalsize Published in \textit{Network Science} 5 (1): 108-139, 2017.\\  Open Access version available at DOI: https://doi.org/10.1017/nws.2017.1}
\tnotetext[t1]{Both authors thank three anonymous reviewers for the \textit{11th International Conference on Web and Internet Economics, WINE 2015, Amsterdam, The Netherlands, December 9-12, 2015}, where this paper was presented.}

\author[adrmatt,secondadrmatt]{Matt V. Leduc\fnref{fn1}}
\ead{mattvleduc@gmail.com}
\fntext[fn1]{Matt V. Leduc wishes to thank Matthew O. Jackson, Ramesh Johari, Peter Glynn, Adrien Vigier, Elena Rovenskaya, Alexander Tarasyev and Stefan Thurner, Brian Fath, Yurii Yermoliev, Tatiana Ermolieva. He is grateful to the National Academy of Sciences, the U.S. National Member Organization for the International Institute for Applied Systems Analysis (IIASA), for support through a grant from the U.S. National Science Foundation during IIASA's YSSP 2013 program.}
\author[insead]{Ruslan Momot\fnref{fn2}}
\ead{ruslan.momot@insead.edu}
\fntext[fn2]{Ruslan Momot thanks Ehud Lehrer (Tel-Aviv University) and Timothy Van Zandt (INSEAD) for invaluable discussions and feedback on the paper. He also wishes to thank Ekaterina Astashkina (INSEAD), Marat Salikhov (INSEAD) and Dmitriy Knyazev (University of Bonn) for their comments and suggestions.}

\address[adrmatt]{Stanford University, MS\&E, 475 Via Ortega, Stanford, CA 94305-4121, USA}
\address[secondadrmatt]{International Institute for Applied Systems Analysis (IIASA), Schlossplatz 1, A-2361 Laxenburg, Austria}
\address[insead]{INSEAD, Boulevard de Constance, Fontainebleau, France, 77305 \\ \ \\ \emph{\textbf{Draft version: Jan. 2017}} }

\begin{abstract}
We study the incentives that agents have to invest in costly protection against cascading failures in networked systems. Applications include vaccination, computer security and airport security. Agents are connected through a network and can fail either intrinsically or as a result of the failure of a subset of their neighbors. We characterize the equilibrium based on an agent's failure probability and derive conditions under which equilibrium strategies are monotone in degree (i.e. in how connected an agent is on the network). We show that different kinds of applications (e.g. vaccination, malware, airport/EU security) lead to very different equilibrium patterns of investments in protection, with important welfare and risk implications. Our equilibrium concept is flexible enough to allow for comparative statics in terms of network properties and we show that it is also robust to the introduction of global externalities (e.g. price feedback, congestion). 
\end{abstract}

\end{frontmatter}

\newpage
\section{Introduction}
Many systems of interconnected components are exposed to the risk of cascading failures. The latter arises from interdependencies or interlinkage, where the failure of a single entity (or small set of entities) can result in a cascade of failures jeopardizing the whole system. This phenomenon occurs in various kinds of systems. Well-known examples include `black-outs' in power grids, where overload redistribution following the failure of a single component can result in a cascade of failures that ripples through the entire grid (e.g. \cite{RosasCasals}, \cite{Wang}). The internet and computer networks also exhibit this phenomenon---one manifestation being the spread of malware (e.g. \cite{Bolot}, \cite{Balthrop}). Likewise, human populations are exposed to the spread of contagious diseases\footnote{For different applications, such as cascading risk in financial systems, see for example \cite{AcemogluSystemic}, \cite{JacksonGolubElliott}.}.

Studying the incentives to guard against the risk of cascading failures in such interconnected systems has received attention in recent years. In early 2015, a measles epidemic spread across the western part of the United States. It was reported that one of the causes was the unwillingness of parents to vaccinate their children (e.g. \cite{economist_vaccination}, \cite{economist_rand_paul}, \cite{USvaccinationRatesReuters}). Indeed, some people may want to avoid the perceived risks of a vaccine's side effects and free-ride on the ``herd immunity'' provided by the vaccination of other people. This raises the following question: what are the incentives to vaccinate against a contagious disease? The same type of question can be asked about other systems subject to the risk of cascading failures. What are the incentives to invest in computer security solutions to protect against the spread of malware? A recent wave of terror attacks within the European union also illustrates the fact that the EU is an interconnected system of many countries. Each member country is thereby exposed to the decisions of other member countries regarding investments in security and intelligence. Indeed, an attacker entering the EU area can reach any location within it. Likewise, what incentives do airports have to invest in security equipment/personnel? How does the structure of interactions between individuals, computers, airports or countries affect those incentives?

There are mainly two streams of literature studying such strategic decisions in interconnected systems. One focuses on the role played by  the structure through which agents interact (e.g. a network), while the other focuses on modeling different types of attacks on the system (e.g. random attacks, targeted attacks, strategic attacks). 

In the first stream of literature, early work studying games of ``interdependent security'' (e.g.  \cite{IDSHealKunreuther} and \cite{IDSHealKunreutherKearns}) considered a broad set of applications ranging from airline security to supply chain management, but did not yet incorporate a complex network interaction structure. More recent work has studied  heterogeneous interaction structures. For example, \cite{GaleottiRogers} consider the problem of a social planner attempting to eradicate an infection from a population. They consider a simple network consisting of two types of agents interacting with others within and across their respective social groups. They then explore the influence of assortativity on the optimal actions of a decision maker. Other papers, like ours, explore the influence of a networked interaction structure on the agents' strategic decisions in more detail. This includes \cite{lelarge2008local} studying the case of strategic immunization and \cite{cabrales2014risk} exploring the setting of interconnected firms choosing investments in risky projects. More recently, \cite{cerdeiro2015contagion} explored the problem of designing the network topology that provides the proper incentives to the agents. 

In the second stream of literature, papers like \cite{GoyalDziubinski} and  \cite{acemoglu2013network} explore strategic attack models, in which a defender chooses protection levels, while an attacker chooses the targets in an attempt to maximize the number of affected agents in the network.

In this paper, we develop a framework to study the incentives that agents have to invest in protection against cascading failures in networked systems. A set of interconnected agents can each fail exogenously (fully randomly) or as a result of a cascade of failures\footnote{Similar random failure mechanisms are studied in \cite{Bolot}, \cite{GoyalVigier}, \cite{aspnes2005inoculation},  \cite{blume2011network} and \cite{acemoglu2013network}.} (through infected connections). Depending on the application, failure can mean a human being contracting an infectious disease, a computer being infected by a virus or an airport/country being exposed to a security event (e.g. a suspicious luggage or passenger being checked in or being in transit). Each agent must decide on whether to make a costly investment in protection against cascading failures. This investment can mean vaccination, investing in computer security solutions or airport security equipment, to name a few important examples. Strategic decisions to invest in protection are based on an agent's intrinsic failure risk as well as on his belief about his neighbors and their probability of failure. In a complex networked system, forming such a belief can be challenging. For that reason, we employ a solution concept that considerably simplifies how agents reason about the network: agents do not observe the network, but simply know the number of connections they have. This is similar to the equilibrium concept used in \cite{Galeotti10}, \cite{JacksonYariv} and \cite{leduc2015pricing}.  This equilibrium concept allows us to preserve the heterogeneity of the networked interaction structure (each agent can have a different degree, i.e. a different number of connections) while simplifying the computation of an equilibrium. It also conveniently allows for comparative statics in terms of the network structure (as captured by the degree distribution), as well as other model parameters. This allows us to measure such things as the effect of an increase in the level of connectedness on investments in protection. 

We characterize the equilibrium for three broad classes of games: (i) \em games of total protection\em, in which agents invest in protection against \em both \em their intrinsic failure risk and the failure risk of their neighbors; (ii) \em games of self protection\em, in which agents invest in protection \em only \em against their intrinsic failure risk; and (iii) \em games of networked-risk protection\em, in which agents invest in protection \em only \em against the failure risk of their neighbors. The first and third classes define games of strategic substitutes, in which some agents free-ride on the protection provided by others. Applications covered by these classes of games include vaccination and standard computer security solutions (e.g. anti-virus). The second class defines a game of strategic complements, in which agents pool their investments in protection and this can result in coordination failures. Applications covered by this class of games include airport security, border security within the European union and other types of computer security solutions (e.g. two-factor authentication (2FA)).

Another of our contributions is to analyze the effect of the network structure on equilibrium behavior in those three classes of games. For example, in the case of vaccination, it is the agents who have \em more \em neighbors than a certain threshold who choose to vaccinate and the agents who are less connected who free-ride. The more connected agents thus bear the burden of vaccination, which can be seen as a positive outcome. In the case of airport security, on the other hand, it is agents who have \em fewer \em neighbors than a certain threshold who choose to invest in security equipment/personnel. Since the less connected airports are less likely to act as hubs that can transmit failures, this can be seen as an inefficient outcome. To our knowledge, we are the first to explicitly characterize such features, which are the consequence of network structure and can have important policy and welfare implications.

Finally, we study the case when the cost of protection is endogenized and allowed to depend on global demand. For instance, the price of vaccines or computer security solutions may increase (e.g. vaccines may be produced in limited supplies) if demand increases. It is important to understand the impact that this may have on agents' behavior as the introduction of such a \em global externality \em (e.g. see the global congestion case in \cite{arribasa2014local}) may conflict with the cascading failure process affecting an agent through his local connections. We characterize the equilibrium after introducing this price feedback and show that the results derived previously still hold with minor changes.

\cite{acemoglu2013network} and \cite{Bolot} are perhaps the closest work to ours. The former paper, in a setting similar to ours, shows that under random and targeted attacks both over- and underinvestment (as compared to the socially optimal level) are possible. Furthermore, the authors show that optimal investment levels are defined by network centrality measures, whereas  our characterization of equilibrium investment is based on degree centrality. Additionally, we further explore the role of the network structure in defining agents' incentives to invest in protection. In particular, we study comparative statics by varying the degree distributions of the underlying network. \cite{Bolot} also consider different types of protection against contagion risk in trees and sparse random graphs. As compared to their probabilistic approach, the equilibrium concept we use allows for a characterization of behavior in terms of an agent's degree. We also deal with a common (possibly endogenized) cost of investment as opposed to their randomized costs. 
Finally, our paper contributes to the rapidly expanding stream of literature on games on networks\footnote{The reader is referred to \cite{JacksonZenouHandbook} for a survey of the existing literature on games on networks.}.

The paper is organized as follows: Section \ref{sec:::cascadingFailures} introduces the concept of cascading failures in networked systems. Section \ref{sec::::MF} develops the game theoretic framework that allows us to study the problem in a tractable way while imposing a realistic cognitive burden on agents. Section \ref{sec::::CharacterizingMFE} characterizes the equilibrium for the three broad classes of games previously mentioned.
Implications for risk and welfare are discussed. Comparative statics results in terms of the network structure (as captured by the degree distribution) and other model parameters are also presented. An extension in which the cost of protection is endogenized is also studied. Section \ref{sec::::conclusion} concludes with a critical evaluation of our model and a discussion of possible extensions. For clarity of exposure, all the proofs are relegated to an appendix.

\section{Cascading Failures in Networks}
\label{sec:::cascadingFailures}
\subsection{Overview}
\begin{figure*}
  \centerline{
\includegraphics[scale=0.8]{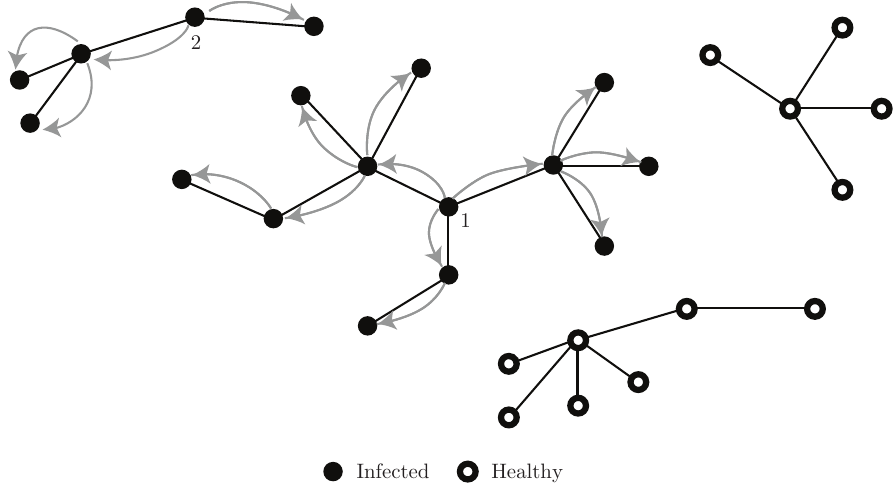}
  }
  \caption{Example of a Contagion Cascade: individuals labeled $1$ and $2$ contract the disease from exogenous sources. From then on, a contagion cascade takes place in discrete steps: all their neighbors become infected. This then leads to their neighbors' neighbors to become infected and so on.}
  \label{fig:::CascadeExample}
\end{figure*}

In this section we will discuss how cascades of failures can propagate through networks. A \em cascade of failures \em is defined as a process involving the subsequent failures of interconnected components. A \em failure \em is a general term that may represent different kinds of costly events. Let us consider, for example, the spread of a disease in a human population. Initially, some individuals get infected through exogenous sources such as livestock, mosquitos or the mutation of a pathogen. These individuals can then transmit the disease through contacts with other humans. Let us suppose that an individual is sure to catch the disease if one of his neighbors is infected. Figure \ref{fig:::CascadeExample} illustrates this process.
We can see the impact of network structure on contagion. Some people lying in certain components remain healthy whereas others are infected by their neighbors. We also see that individuals with a high number of contacts tend to facilitate contagion. This is a simplified model of contagion. A more realistic model could, for example, transmit the disease only to randomly selected neighbors, depending on its virulence. 

\begin{figure*}
\centering
\includegraphics[scale=0.8]{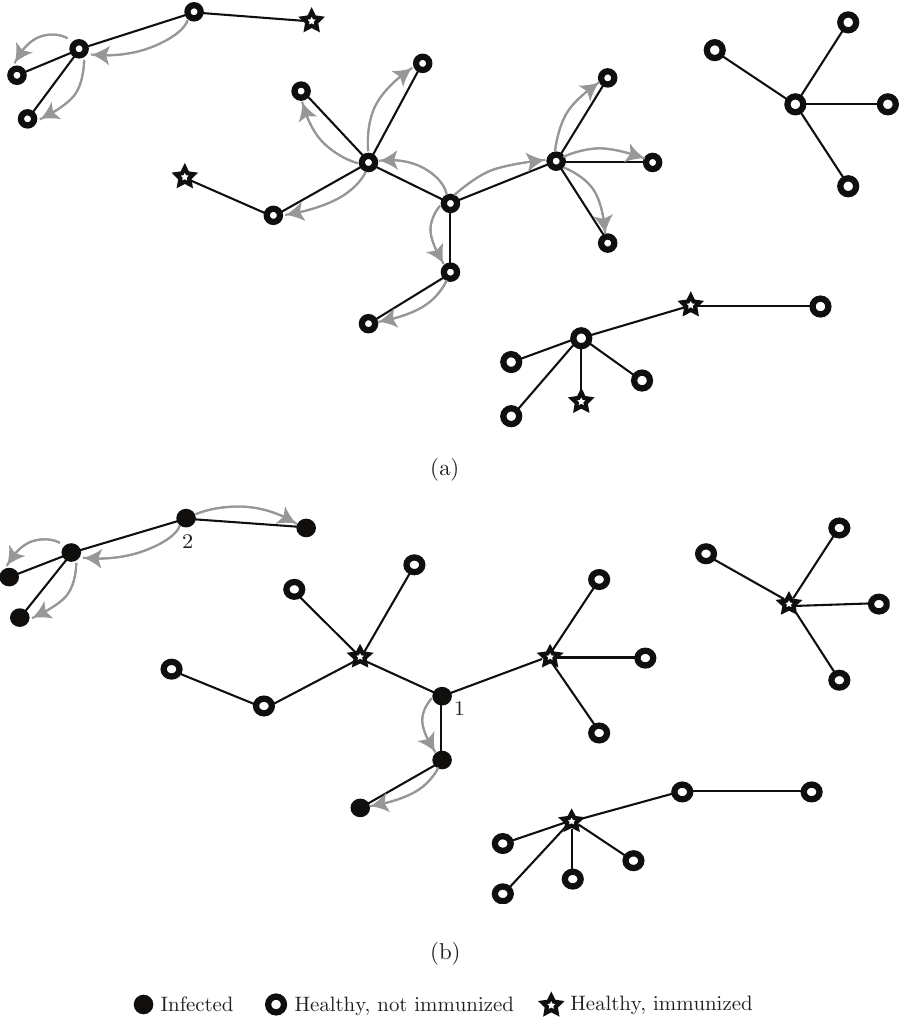}
\caption{Examples of Contagion Cascades in the Presence of Immunized Individuals: individuals labeled $1$ and $2$ contract the disease from exogenous sources. The contagion cascade then propagates. In part (a), a randomly-chosen subset of agents were vaccinated against the disease. In part (b), individuals with at least $4$ contacts were vaccinated against the disease.}
  \label{fig:::Immunization}
\end{figure*}

Now let us imagine that some individuals are vaccinated and therefore are not susceptible to becoming infected, neither by exogenous sources nor by contacts with other people. This will have an impact on the cascading process. Indeed, it will effectively `cut' certain contagion channels, thereby impeding the spread of the disease. Figure \ref{fig:::Immunization} illustrates this. We see that the importance of the network structure becomes even more striking. In Fig.~\ref{fig:::Immunization}a), immunized individuals have been selected randomly, whereas in Fig.~\ref{fig:::Immunization}b) individuals with $4$ or more contacts have been immunized. It is clear that those more connected individuals often act as hubs through which contagion can spread more easily. When these individuals are immunized, the effect of impeding the propagation of the disease tends to be much greater than when the immunized individuals are chosen at random.

In this example, the `failure' of an individual means he becomes infected by the disease. In other applications, `failure' can mean infection by malware. The nodes then no longer represent individuals but computers (or local subnetworks or autonomous systems). Antivirus software or other sorts of computer security solutions are means by which the spread of malware can be impeded.

We saw in the simple example of Fig.~\ref{fig:::Immunization} that the configuration of the vaccinated nodes was crucial to impeding contagion. An important question is to study the incentives that an individual may have to become vaccinated. How does the network structure affect his decision to become vaccinated? What roles other individuals play in influencing that decision through their own vaccination behavior?

Given the range of applications, we will talk of an \em investment in protection\em. This refers to an investment made by a node in order to protect itself against the risk of failure. In the next section, we build a model of strategic investment in protection against cascading failures in networked systems. We will refer to nodes as \em agents\em, since they make decisions regarding this investment in protection. More generally, we will be interested in how the network structure and the failure propagation mechanism influence those decisions through the externalities that they generate.

\subsection{Network}
\label{sec:::Model}

A network, as the one described previously, can be formally defined as follows. There is a set of nodes (or agents) $\mathcal{N} = \{1,2, ..., n\}$. The connections between them are described by an \textit{undirected} network that is represented by a symmetrical adjacency matrix $g \in \{0,1\}^{n \times n}$, with $g_{ij}=1$ implying that $i$ and $j$ are connected. $i$ can thus be affected by the failure of $j$ and vice versa. By convention, we set $g_{ii}=0$ for all $i \in \mc{N}$. The network realization $g$ is drawn from the probability measure $P:  {\{0,1 \}}^{n \times n} \rightarrow [0,1]$ over the set of all possible networks with $n$ nodes. We assume that $P$ is permutation-invariant, i.e. that changing node labels does not change the measure. Each agent $i$ has a neighborhood $N_i(g) = \{ j | g_{ij} = 1 \}$. The \em degree \em of agent $i$, $d_i(g)$, is the number of $i$'s connections, i.e. $d_i(g) = | N_i(g) | $.

\section{A Bayesian Network Security Game}
\label{sec::::MF}

\subsection{Informational Environment}

We study an informational environment similar to the one presented in \cite{Galeotti10}. Agents are aware of their proclivity to interact with others, but do not know who these others will be when taking actions. Formally, this means that an agent knows only his degree $d_i$. For example, a bank may have a good idea of the number of financial counter-parties it has but not the number of counter-parties the latter have, let alone the whole topology of the interbank system. In applications to the spread of contagious diseases, an individual may know the number of people he interacts with, but not the number of people the latter interact with. Likewise, in the case of an email network, someone may know the number of contacts he has, but not the number of contacts his contacts have. 

First, since $P$ is permutation invariant (cf. Section \ref{sec:::Model}), we can define the {\em degree distribution} of $P$ as the probability a node has degree $d$ in a graph drawn according to $P$; we denote the degree distribution\footnote{Throughout, we use the term \textit{degree distribution} to mean \textit{degree density}. When referring to the \textit{cumulative distribution function (CDF)}, we will do so explicitly.} by $f(d)$ for $d \geq 1$. Note that we are not interested in modeling agents of degree $0$ (since they do not play a game) and we therefore always assume that $f(0)=0$. We assume a countably infinite set of agents. An agent's type is his degree $d$ and it is drawn i.i.d. according to the degree distribution $f(d)$. Likewise, the degree of each of an agent's neighbors is drawn i.i.d. according to the density function $\tilde{f}(d)$. This is the \em edge-perspective degree distribution \em and can be written as $\tilde{f}(d) = \frac{f(d) d}{\sum_{d' \geq 1} f(d') d'}$. This expression follows from a standard calculation in graph theory (see \cite{JacksonBook} for more details). $\tilde{f}(d)$ is the probability that a neighbor has degree $d$. It therefore takes into account the fact that a higher-degree node has a higher chance of being connected to any agent and thus of being his neighbor. Thus agents reason about the graph structure in a simple way through the degree distribution.

\subsection{Action Sets and Strategies}

In order to protect himself against the risk of failure, we allow an agent $i$ to make a costly investment in \em protection\em. This is a \textit{one-shot investment} that can be made \em in anticipation \em of a cascade of failures, which may take place in the future. This investment in protection is represented by an action $a_i$, which is part of a binary action set $\mathcal{A} = \{0,1\}$. The latter represents the set of possible investments in protection against failure: $a_i=1$ means that the agent  invests in protection while $a_i=0$ means that the agent remains unprotected. In an application to computer security, $a_i$ can represent an investment in computer security solutions or anti-virus software. In applications to disease spread, $a_i$ can represent vaccination, whereas in the case of airport security, $a_i$ can represent an investment in security personnel or equipment. We assume throughout that $\mc{A}$ is the same for all agents. The exact effect of this action on an agent's actual failure risk will be formalized later in Definition \ref{def:::EffFailureProb}.

Note that all agents have access to the same information about the network (only its degree distribution $f(d)$). An agent does not know his position in the network, only the number of neighbors he has (an agent's degree is his type). An agent $i$'s behavior is thus governed only by his degree $d_i$ and not by his label $i$. We can then define a strategy in the following way. 

\begin{definition}
A strategy $\mu: \mathbbm{N}^+ \rightarrow [0,1]$ is a scalar-valued function that specifies, for every $d >0 $, the probability that an agent of degree $d$ invests in protection. We denote by $\mc{M}$ the set of all strategies.
\end{definition}

Thus $\mu(d)$ is the symmetric mixed strategy played by an agent of degree $d$. Note that $\mc{M} = [0,1]^{\infty}$, the space of $[0,1]$-valued sequences. Throughout, we endow $\mc{M}$ with the product topology and $[0,1]$ with the Euclidean topology.

\subsection{Failure Probabilities and Utility Functions}
\label{sec:FailProbs_utilities}

We start with the following definition:

\begin{definition}
An agent's intrinsic failure probability is denoted by $p \in [0,1]$.
\end{definition}

We thus assume all agents can fail intrinsically with the same probability $p$. The interpretation of intrinsic failure depends on the application. In the context of malware, intrinsic failure means a computer becoming infected as a result of a direct hacking attack. In the context of the spread of contagious diseases, intrinsic failure means being infected by a virus through non-human sources, such as contact with livestock or insects. In the context of airport security, intrinsic failure can mean a suspicious luggage being checked in at the airport. 

We now state a property of this network security game, which addresses how an agent reasons about the failure probability of his neighbors.
\begin{property}
\label{mfa::::adopt}
Each agent conjectures that each of his neighbors fails with probability $\mc{T}(\mu) \in [0,1]$, independently across neighbors.
\end{property}

This setting is similar to that of \cite{JacksonYariv}, where each neighbor adopts a product or an opinion with a certain probability that depends on the strategy played by the population. Note that the dependence of a neighbor's failure probability $\mc{T}(\mu)$ on the strategy $\mu$ played by other agents was made explicit. An agent's cascading failure probability can now be defined in terms of $\mc{T}(\mu)$, as seen in the following definition.

\begin{definition}
\label{def:qd}
For any $d$, let the function $q_d: [0,1] \rightarrow [0,1]$ denote a degree-$d$ agent's cascading failure probability, i.e. $q_d(\mc{T}(\mu))$ is the probability that an agent of degree $d$ will fail as a result of a cascade of failures, given that his neighbors each fail independently with probability $\mc{T}(\mu)$.
For any $d$, $q_d(\mc{T}(\mu))$ is strictly increasing and continuous in $\mc{T}(\mu)$. Moreover, we explicitly set $q_0(\mc{T}(\mu)) = 0$ and thus an agent with no neighbor cannot fail as a result of a cascade of failures.
\end{definition}

The actual expression for $q_d(\mc{T}(\mu))$ depends on the type of cascade we are considering. We will consider only a situation where $\{q_d\}_d$ is an increasing sequence of functions. That is, when $d' > d$, then $q_{d'}(\mc{T}(\mu)) > q_{d}(\mc{T}(\mu)) $ for any $\mc{T}(\mu) \in [0,1]$. In other words, the cascading failure risk is higher when an agent has more connections\footnote{The reader is referred to Chapter $4$ of \cite{MLeducThesis} for the case where $q_d(\mc{T}(\mu))$ is decreasing in $d$. This can model a form of diversification of failure risk across neighbors.}. For convenience, we will sometimes write $q_d(\mc{T}(\mu))$ simply as $q_d$. 

Since an agent of degree $d$ either fails intrinsically with probability $p$ or in a cascade with probability $q_d$, we can define his \em total probability of failure \em as follows.

\begin{definition}[Total probability of failure]
The total probability of failure of an agent of degree $d$ is
 \begin{equation}
 \label{def:::totalFailureProb}
\beta_d = p + (1-p) q_d.
\end{equation}
\end{definition}

Thus an agent can either fail intrinsically (i.e. by himself) or as a result of the failures of a subset of his neighbors. Those neighbors who have failed may have done so intrinsically or as a result of the failure of a subset of their own neighbors.

We study a static setting, in which agents make decisions simultaneously, \em in anticipation \em of a cascade of failures that may happen in the future. Therefore each agent is \em healthy \em when he chooses an action $a \in \mathcal{A}$ representing a costly investment in protection against failure. This is a good decision model for the applications that we cover. E.g., vaccines are taken by healthy individuals in anticipation of an epidemic that may spread in the future. Likewise, investments in computer security solutions are taken for healthy computers or autonomous systems in anticipation of the spread of malware, which may take place at a later date. Similar long-term security decisions are taken in other contexts, such as airport security, for example.

We now describe how this action affects an agent's failure probability.

\begin{definition}
\label{def:::EffFailureProb}
Let the mapping $\mc{B}:[0,1] \times [0,1] \times \mathcal{A} \rightarrow [0,1]$ denote the effective failure probability of an agent. We assume that $\mc{B}(p,q_d,a)$ is continuous in all arguments, increasing in $p$ and in $q_d$ and that it is decreasing in $a$.
\end{definition}

Thus, $\mc{B}(p,q_d,a)$ is the total failure probability of an agent (defined in (\ref{def:::totalFailureProb})) when he has invested $a$ in protection against failure. Note that this definition allows this action to operate separately on $p$ and $q_d$, as will be seen in Section \ref{sec::::CharacterizingMFE}. This will become useful as we study different kinds of protection. We can now state an agent's expected utility function, which will capture his decision problem.

A degree-$d$ agent's expected utility function is given by
\begin{equation}
\label{eq::::U}
U_d(a,\mu) = -V \cdot \mc{B} (p,q_d(\mc{T}(\mu)),a) - C \cdot a.
\end{equation}
where $C>0$ is the cost of investing in protection, $V>0$ is the value that is lost in the event of failure and $\mc{B}(\cdot,\cdot,\cdot)$ is the \em  effective failure probability \em (cf. Definition \ref{def:::EffFailureProb}).

This utility function captures the tradeoff between the expected loss $\let\ab\allowbreak V \cdot \mc{B} (p, q_d(\mc{T}(\mu)), a) $ and the cost\footnote{The cost of investing in protection may represent the price of airport security equipment or computer security solutions. It may also represent the possible side-effects that may be associated with a vaccine (e.g.  \cite{economist_rand_paul}).} $C$ of investing in protection. Notice again that an agent's expected utility depends on the actions of others only through the cascading failure probability $q_d(\mc{T}(\mu))$, since they will affect the probability of failure $\mc{T}(\mu)$ of a randomly-picked neighbor. Note also that the expected utility function\footnote{Note that we could write a degree-$d$ agent's expected utility function as $U(a,\mu,d)$. We write it with $d$ as a subscript simply because it is  a convenient notation.} $U_d(\cdot,\cdot)$ depends on the agent's degree $d$ but not on his identity $i$. Therefore, any two agents $i$ and $j$ who have the same degree have the same expected utility function. From the assumptions on $\mc{B}$, $U_d$ is continuous in all arguments. An agent is risk-neutral and will thus maximize this expected utility function by choosing the appropriate action $a$. The game thus models security decisions under contagious random attacks in a network where each agent (node) knows only his own degree and the probability that a neighbor has a certain degree. 

While the cascading failure probability $q_d$ can take many functional forms, we provide several examples which can all be modeled using the particular form $q_d(\mathcal{T}(\mu)) = 1-(1-r\mathcal{T}(\mu))^d$. This functional form results from a contact process.\\[0.1cm]

\noindent\textbf{Malware or Virus Spread:}
Let a computer be infected by a direct hacking attack with probability $p$. Assume that malware (i.e. computer viruses) can spread from computer to computer according to a general contact process: if a neighbor is infected, then the computer will be infected with probability $r$. If each neighbor is infected with probability $\mathcal{T}(\mu)$ and this infection spreads independently across each edge with probability $r$, then $q_d(\mathcal{T}(\mu)) = 1-(1-r\mathcal{T}(\mu))^d$. This contact process can also serve as a model for the spread of viruses among human populations. In this case, $p$ is the probability of being infected by non-human sources (e.g. insects, livestock, etc.) and $q_d(\mathcal{T}(\mu))$ is the probability of being infected by neighbors (i.e. other persons with whom the agent interacts). The parameter $r$ models the virulence or infectiousness of the process: given that a neighbor is infected, $r$ is the probability\footnote{In Fig. \ref{fig:::CascadeExample} and Fig. \ref{fig:::Immunization}, $r$ was assumed to be $1$ for simplicity of exposure.} that he will infect the agent.\\[0.1cm]

\noindent\textbf{Airport and European Union Security:}
The contact process described above can also be applied to airport or EU security. The exogenous failure (with probability $p$) can mean a security event such as the failure to stop a suspicious luggage from being checked in on a flight or a terrorist entering the European union from outside through one of the EU countries with weaker border control. In these scenarios, the agents represent airports or countries, and the edges linking them represent flights and connecting routes between countries. The suspicious luggage or terrorist can then cascade, i.e. travel to one or more other airports/countries, exposing them to security risks. $q_d(\mathcal{T}(\mu)) = 1-(1-r\mathcal{T}(\mu))^d$ can then model the risk of an entity coming into contact with a security threat coming from a neighboring country or airport. \\[0.1cm]

In the next two sections we develop both the optimal response of an agent to the environment described previously, as well as the consistency check that $\mc{T}(\mu)$ should satisfy given the strategic choices of the agents.

\subsection{Consistency}
We will now develop a consistency check that guarantees that a randomly-picked neighbor's failure probability $\mc{T}(\mu)$ is consistent with the strategy $\mu$ played by the population.

\begin{definition}
\label{def::::Tgeneral}
Let the function $\mc{F}: \mc{M} \times [0,1] \rightarrow [0,1]$ be defined as
\begin{equation}
\label{eq::::Fgeneral}
\mathcal{F}(\mu,\alpha) = \sum_{d \geq 1} \tilde{f}(d) \mc{B}(p,q_{d-1}(\alpha),\mu(d)). 
\end{equation}
\end{definition}

In the above definition\footnote{Note that an agent does not internalize the effect of his own failure on others when forming his belief about the failure risk of a neighbor. Hence the presence of $q_{d-1}(\alpha)$ on the right-hand side of (\ref{eq::::Fgeneral}) instead of $q_{d}(\alpha)$: the cascading failure risk of a given neighbor of degree $d$ is only due to his $d-1$ other neighbors.}, $\mc{F}(\mu,\alpha)$ is the failure probability of a randomly-picked neighbor given that agents play strategy $\mu$ and this neighbor's other neighbors fail with probability $\alpha$. A fixed point $\alpha = \mc{F}(\mu,\alpha)$ ensures that $\alpha$ is the same across all agents and consistent with $\mu$. We consider $\mc{F}(\mu,\alpha)$ with the following property:
\begin{property}
\label{ass::::T_unique}
For any $\mu \in \mc{M}$, $\mc{F}(\mu,\alpha)$ has a unique fixed point in $\alpha$.
\end{property}


Note that Property \ref{ass::::T_unique} is not particularly stringent. It is easy to verify in the contact process models of the examples described in Section \ref{sec:FailProbs_utilities}. 

We can now formally define $\mc{T}(\mu)$, the failure probability of a randomly-picked neighbor given that strategy $\mu$ is played by other agents:

\begin{definition}
 Given $\mc{F}: \mc{M} \times [0,1] \rightarrow [0,1]$ satisfying Property \ref{ass::::T_unique}, let $\mathcal{T}: \mc{M}  \rightarrow [0,1]$ be defined as follows: for any $\mu \in \mc{M}$,
\begin{equation}
\label{eq::::Tgeneral}
\mc{T}(\mu) = \mc{F}(\mu,\mc{T}(\mu)).
\end{equation}
\end{definition}


\subsection{Optimal Response}
It is now straightforward to solve for the optimal strategy of an agent of degree $d$: an agent invests in protection, does not invest, or is indifferent if $U_d(1,\mu)$ is greater than, less than, or equal to  $U_d(0,\mu)$, respectively.  We thus have the following definition.

\begin{definition}
\label{def::::S}
Let $\mc{S}_d(\mc{T}(\mu)) \subset [0,1]$ denote the set of optimal responses for a degree-$d$ agent given $\mc{T}(\mu)$; i.e.:
\begin{align*}
U_d(1,\mu) > U_d(0,\mu) &\implies \mc{S}_d(\mc{T}(\mu)) = \{1\};\\
U_d(1,\mu) < U_d(0,\mu)&\implies \mc{S}_d(\mc{T}(\mu)) = \{0\};\\
U_d(1,\mu) = U_d(0,\mu) &\implies \mc{S}_d(\mc{T}(\mu)) = [0,1].
\end{align*}
We can now let $\mc{S}(\mc{T}(\mu)) \subset \mc{M}$ denote the set of optimal strategies given $\mc{T}(\mu)$; i.e.,
\[ \mc{S}(\mc{T}(\mu)) = \prod_{d \geq 1} \mc{S}_d(\mc{T}(\mu)). \]
\end{definition}

Note that at least one optimal response always exists and is essentially uniquely defined, except at those degrees where an agent is indifferent.

\subsection{Equilibrium}
\label{sec::::MFE}
We now formally define the equilibrium concept and state our first proposition.

\begin{definition}[Mean-Field Equilibrium]
A strategy $\mu^*$ constitutes a mean-field equilibrium (MFE) if $\mu^* \in \mathcal{S}(\mathcal{T}(\mu^*))$.
\end{definition}

This equilibrium definition ensures that both the optimality and consistency conditions are satisfied. Also note that to any equilibrium $\mu^*$, there corresponds a unique equilibrium neighbor failure probability $\alpha^* = \mathcal{T}(\mu^*)$.

\begin{proposition}[Existence]
\label{th::::existence}
Any network security game that satisfies Properties \ref{mfa::::adopt} and \ref{ass::::T_unique} has a mean-field equilibrium.
\end{proposition}

An MFE is a symmetric equilibrium with the property that an agent's neighbors fail independently with the same probability $\mc{T}(\mu^*)$ under $\mu^*$.
An MFE is particularly easy to compute. In fact, $\alpha^*=\mc{T}(\mu^*)$ is obtained from a one-dimensional fixed-point equation resulting from the composition of $\mc{T}$ and $\mc{S}$, i.e. $\alpha^* = \mc{T}(\mc{S}(\alpha^*))$. $\mu^*$ is then found from the map $\mc{S}(\alpha^*)$ (cf. Definition \ref{def::::S}). Allowing for correlations between the failures of neighbors would considerably complicate the analysis\footnote{For some work in that direction, see Chapter 3 of \cite{MLeducThesis}.}.
\section{Characterizing Equilibria}
\label{sec::::CharacterizingMFE}

In this section, we will study three classes of games in which agents make decisions to invest in protection. We will start with games of \em total protection\em, in which an agent's investment decreases his total risk of failure. We will then proceed with games of \em self protection\em, in which an agent's investment in protection only protects him against his own intrinsic risk of failure. We will finally study an intermediate case: a game of \em networked-risk protection\em, in which an agent's investment in protection only protects him against the risk of failure of his neighbors 

\subsection{Games of Total Protection}
\label{sec::::totProt}

In games of \em total protection\em, the investment protects both against the intrinsic failure risk and the cascading failure risk. 

Examples of applications covered by this class are the spread of contagious diseases and the decision to vaccinate or malware and the investment in anti-virus or computer security solutions. Vaccination, for example, protects against both the risk of being infected by non-human (intrinsic failure risk) and human sources (cascading failure risk). It is also the case for standard anti-virus software featuring a firewall protection. This protects an agent against both direct hacking attacks (intrinsic failure risk) and malware spread through the Internet/e-mail networks (cascading failure risk). 

We have the following definition:

\begin{definition}[Games of total protection]
\label{def::::totalProt}
In a game of total protection, the effective failure probability has the following form
\begin{equation}
\label{eq::::BtotalProt}
\mc{B}(p,q_d(\mc{T}(\mu)),a) = \Big(p + (1-p) q_d(\mc{T}(\mu)) \Big) \cdot (1 - k a)
\end{equation}
for some $k \in [0,1]$ and 
\begin{equation}
\label{eq::::FtotalProt}
\mathcal{F}(\mu,\alpha) = \sum_{d \geq 1} \tilde{f}(d) \Big(p + (1-p) q_{d-1}(\alpha)  \Big) \cdot (1 - k \mu(d)). 
\end{equation}
\end{definition}
In games of total protection, as can be seen in (\ref{eq::::BtotalProt}), an agent's investment in protection decreases his total probability of failure $p + (1-p) q_d(\mc{T}(\mu))$. The parameter $k$ governs the effectiveness of the investment in protection. The higher $k$, the more an investment in protection reduces the failure probability.

Before stating our first theorem, we introduce the following definition.

\begin{definition}[Upper-threshold strategy]
\label{def::::upperthreshold}
A strategy $\mu$ is an upper-threshold strategy if there exists $d_U \in \mathbbm{N}^+ \bigcup \{ \infty \}$, such that:
\begin{align*}
d < d_U &\implies \mu(d) = 0;\\
d > d_U &\implies \mu(d) = 1.
\end{align*}
\end{definition}

Thus, under an upper-threshold strategy, agents with degrees \em above \em a certain threshold invest in protection whereas agents with degrees \em below \em that threshold do not invest. Note that the definition above does not place any restriction on the strategy {\em at} the threshold $d_U$ itself; we allow randomization at this threshold. 

Games of total protection are \em submodular\em. In other words, they are of \em strategic substitutes\em: the more other agents invest in protection (the lower $\mc{T}(\mu)$), the less an agent has an incentive to invest in protection. A nice property of games of total protection is that they have a unique equilibrium that is characterized by an upper-threshold strategy.

\begin{theorem}[Total Protection]
\label{th::::uniqueness}
In a game of total protection, the equilibrium $\mu^*$ is unique. Moreover, $\mu^*$ is an upper-threshold equilibrium, i.e. $\mu^*$ is an upper-threshold strategy.
\end{theorem}

The intuition behind this result is that, higher-degree agents are more exposed to cascading failures than lower-degree agents, thus making an investment in \em total \em protection relatively more rewarding. The implications of this theorem are important as higher-degree agents are more likely to act as hubs though which contagion can spread. This result can thus be seen as a satisfactory outcome since more connected agents have higher incentives to internalize the risk they impose on the system. In equilibrium, the total cost of protection is thus born by those who have a maximal effect on decreasing $\mathcal{T}(\mu)$. For example, in the case of malware, agents with a higher level of interaction (higher degree) have a higher incentive to invest in computer security (i.e. anti-virus software). The same principle applies in the case of human-born viruses: individuals who interact more have a higher incentive to get vaccinated.

Note that in spite of the above, agents tend to underinvest in equilibrium compared to the socially optimal investment level. This is the result of free-riding and is in line with classical results of moral hazard in economics and the failure of agents to take into account negative externalities. 

In the next section, we study the second class of games: Games of \em self protection\em.

\subsection{Games of Self Protection}
\label{sec::::GamesSelfProt}

In games of \em self protection\em, the investment protects only against the intrinsic failure risk. 

Examples of applications covered by this class of games include airport security when luggage/passengers are only scanned at the originating airport. Airports then otherwise rely on each other's provision of security for transiting passengers/luggage. The same principle applies to security within the European Union, where travelers are only inspected at their point of entry. EU countries otherwise rely on each other's security for travelers within the EU. 

Another important example is two-factor authentication (2FA) in computer networks. Consider an e-mail network and a provider such as Gmail. The latter allows its users to use such a two-factor authentication (2FA) feature. Users who take advantage of this option are asked to enter a security code sent to their mobile phone in addition to their password entered upon authentication. 2FA thus effectively protects against direct hacking attacks (a user's personal intrinsic risk). Indeed, access to the account with 2FA can only be granted conditional on the user having access to the mobile phone linked to this account. Yet, 2FA does not diminish the user's exposure to cascading failure risk (i.e. malware transmitted through the e-mail network): carelessly opening an infected e-mail attachment from a friend can fully compromise his account. 

We now have the following definition:

\begin{definition}[Games of self protection]
\label{def::::selfProt}
In a game of self protection, the effective failure probability has the following form
\begin{equation}
\label{eq::::BselfProt}
\mc{B}(p,q_d(\mc{T}(\mu)),a) = p \cdot (1-ka) + (1-p \cdot (1-ka)  ) \cdot q_d(\mc{T}(\mu)) 
\end{equation}
for some $k \in [0,1]$ and 
\begin{equation}
\label{eq::::FselfProt}
\mathcal{F}(\mu,\alpha) = \sum_{d \geq 1} \tilde{f}(d) \Big( p \cdot (1 - k \mu(d)) + \big(1-p \cdot (1 - k \mu(d))  \big) \cdot q_{d-1}(\alpha) \Big). 
\end{equation}
\end{definition}

In games of self protection, as can be seen in (\ref{eq::::BselfProt}), an agent's investment in protection only decreases his intrinsic probability of failure $p$. It has no effect on his cascading failure probability $q_d(\mc{T}(\mu))$. Again, the parameter $k$ governs the effectiveness of the investment in protection corresponding to the action $a$.

Before stating our second theorem, we introduce the following definition.

\begin{definition}[Lower-threshold strategy]
\label{def::::lowerthreshold}
A strategy $\mu$ is a lower-threshold strategy if there exists $d_L \in \mathbbm{N}^+ \bigcup \{ \infty \}$, such that:
\begin{align*}
d > d_L &\implies \mu(d) = 0;\\
d < d_L &\implies \mu(d) = 1.
\end{align*}
\end{definition}

Under a lower-threshold strategy, agents with degrees \em below \em a certain threshold invest in protection whereas agents with degrees \em above \em that threshold do not invest. Note that the definition above does not place any restriction on the strategy {\em at} the threshold $d_L$ itself; we allow randomization at this threshold. 

Games of \em self protection \em are \em supermodular\em. In other words, they are of \em strategic complements\em: the more other agents invest in protection (the lower $\mc{T}(\mu)$), the more an agent has an incentive to invest in protection. Since games of self protection are effectively coordination games, there can be multiple equilibria. The next result shows that any equilibrium can be characterized by a lower-threshold strategy. In other words, the thresholds are reversed when compared to games total protection (cf. Theorem \ref{th::::uniqueness}).

\begin{theorem}[Self Protection]
\label{th::::thresholdSelfProtection}
In a game of self protection, any equilibrium $\mu^*$ is a lower-threshold equilibrium. That is, $\mu^*$ is a lower-threshold strategy.
\end{theorem}

\em Higher \em cascade risk thus leads to \em lower \em incentives to invest in protection. This is because an agent remains exposed to the failure risk of others irrespectively of whether he invests in protection. An investment in protection thus has lower returns as the cascading failure risk increases. An agent's cascading failure risk \em increases \em in degree, and thus higher-degree agents invest \em less \em in protection than lower-degree agents. The intuition is that higher-degree agents are more exposed to cascading failure risk than lower-degree agents, thus making an investment in their own \em self \em protection relatively less rewarding. 

In the example of airport security, an airport that interacts with a high number of other airports has smaller incentives to invest in its own security, since it remains exposed to a high risk of being hit by an event coming from a connecting flight. This, as before, is assuming that the passengers/luggage are only inspected at their point of origin and not at points of transit. In the example of two-factor authentication (2FA) in an email network, it is the users with a high number of contacts who have lower incentives to enable this security feature since they remain exposed to infected email attachments from their many contacts.   

The fact that, in games of self-protection, the incentives are reversed  has important implications. In fact, the more connected (higher-degree) agents have a lesser incentive to invest in protection even though they are more vulnerable \em and \em more dangerous, i.e. they are hubs through which cascading failures can spread. More central agents thus have lower incentives to internalize the risk they impose on the system, pointing to an inefficient outcome. Moreover, in equilibrium, the total cost of protection is born by lower-degree agents: those who have the smallest effect on decreasing $\mathcal{T}(\mu)$.

\subsection{Games of Networked-Risk Protection}
\label{sec:::networkedriskprot}

In games of \em networked-risk protection\em, the investment protects only against the cascading failure risk. It does not protect against intrinsic failure risk. 

Examples of applications include protection against many sexually transmitted diseases. For instance, the use of condoms protects against the transmission of HIV/AIDS through sexual partners. Nevertheless, such practices leave agents exposed to the external risk of being infected through a medical mistake in a hospital (e.g. with an infected syringe). 

We have the following definition:

\begin{definition}[Games of networked-risk protection]
\label{def::::netProt}
In a game of networked-risk protection, the effective failure probability has the following form
\begin{equation}
\label{eq::::BnetProt}
\mc{B}(p,q_d(\mc{T}(\mu)),a) = p + (1-p) \cdot q_d(\mc{T}(\mu)) \cdot (1-ka)
\end{equation}
for some $k \in [0,1]$ and 
\begin{equation}
\label{eq::::FnetProt}
\mathcal{F}(\mu,\alpha) = \sum_{d \geq 1} \tilde{f}(d) \Big( p + \big(1-p \big) \cdot q_{d-1}(\alpha) \cdot (1 - k \mu(d))  \Big). 
\end{equation}
\end{definition}
We now show that a game of networked-risk protection is structurally equivalent to a game of total protection.
\begin{corollary}
\label{cor:networkedrisk}
A game of networked-risk protection is structurally equivalent to a game of total protection. Particularly, an equilibrium strategy $\mu^*$ in any game of networked-risk protection is unique and is characterized by an upper threshold. 
\end{corollary}

It is easy to see that agents have lesser incentives to invest than in the case of a game of total protection. Indeed, the marginal utility of investing in the latter case is always $Vpk$ higher, because an investment also protects against the intrinsic failure risk. We thus conclude that $\mu^*_{tp} \succeq \mu^*_{np}$, where $\mu^*_{tp}$ and $\mu^*_{np}$ are the investment profiles in games of total and networked-risk protection, respectively. In other words, if an agent of some degree invests in the case of networked-risk protection, then he will necessarily also invest in the case of total protection. In the interest of space, we skip further in-depth discussion of the results in this section as they mainly replicate the results of Section \ref{sec::::totProt}.
\subsection{Welfare, Risk and Comparative Statics}

The next proposition states when the equilibrium  expected utility and effective failure risk of an agent are monotone in degree.

\begin{proposition}[Risk and Welfare I]
\label{prop:risk_welfare_I}
Let $a_d \in \mu^*(d)$:
\begin{itemize}

\item{(i)} The equilibrium expected utility $U_d(a_d,\mu^*)$ is non-increasing in $d$. 

\item{(ii)}  In a game of self protection, the equilibrium effective failure probability\\ $\let\ab\allowbreak \mc{B}(p,q_d(\mc{T}(\mu^*)), a_d)$ is non-decreasing in $d$. 

\end{itemize}
\end{proposition}

Note that there is no analogue to Part (ii) for games of total protection or networked-risk protection. The equilibrium effective failure probability can be non-monotone in degree. Indeed, the upper-threshold strategy means that higher-degree agents invest in protection and may thus have a lower effective failure probability than lower-degree agents.

We will now state a welfare result for games of self protection. These games are easier to analyze because they are of strategic complements. In games of self-protection, agents effectively \em pool \em their investments in protection and, as said earlier, there can be multiple equilibria. These equilibria can however be ordered by level of investment. Suppose there are $m$ possible equilibria. Then, they can be ordered in the following way
\begin{equation*}
\mu^*_1 \preceq \mu^*_2 \preceq ... \preceq \mu^*_m.
\end{equation*} 
Since (\ref{eq::::FselfProt}) is decreasing in $\mu$, it follows that $ \mc{T}(\mu^*_1)  \geq \mc{T}(\mu^*_2) ... \geq \mc{T}(\mu^*_m)$.

We then have a second welfare result.

\begin{proposition}[Welfare II]
\label{prop::::welfare_II}
In a game of self protection, let $\mu^*_k \preceq \mu^*_l$ be two equilibria ordered by level of investment. Then $\mu^*_l$ weakly Pareto-dominates $\mu^*_k$.
\end{proposition}

This result is not trivial. It effectively states that in the high-investment equilibrium, the decrease in risk resulting from higher investments outweighs the cost of those investments. This is due to the positive externality stemming from the effect of pooled investments in protection, which reduce all agents' failure risk.
  
We can focus our attention on the minimum-investment equilibrium $\underline{\mu}^*$ and the maximum-investment equilibrium $\bar{\mu}^*$. In the former, $\mathcal{T}(\underline{\mu}^*)$ is actually maximal since agents invest least, while in the latter, $\mathcal{T}(\bar{\mu}^*)$ is actually minimal since agents invest most. From Proposition \ref{prop::::welfare_II}, agents playing the minimum-investment equilibrium can be thus considered a coordination failure.

\begin{figure*}
\centering
  \includegraphics[scale=0.8]{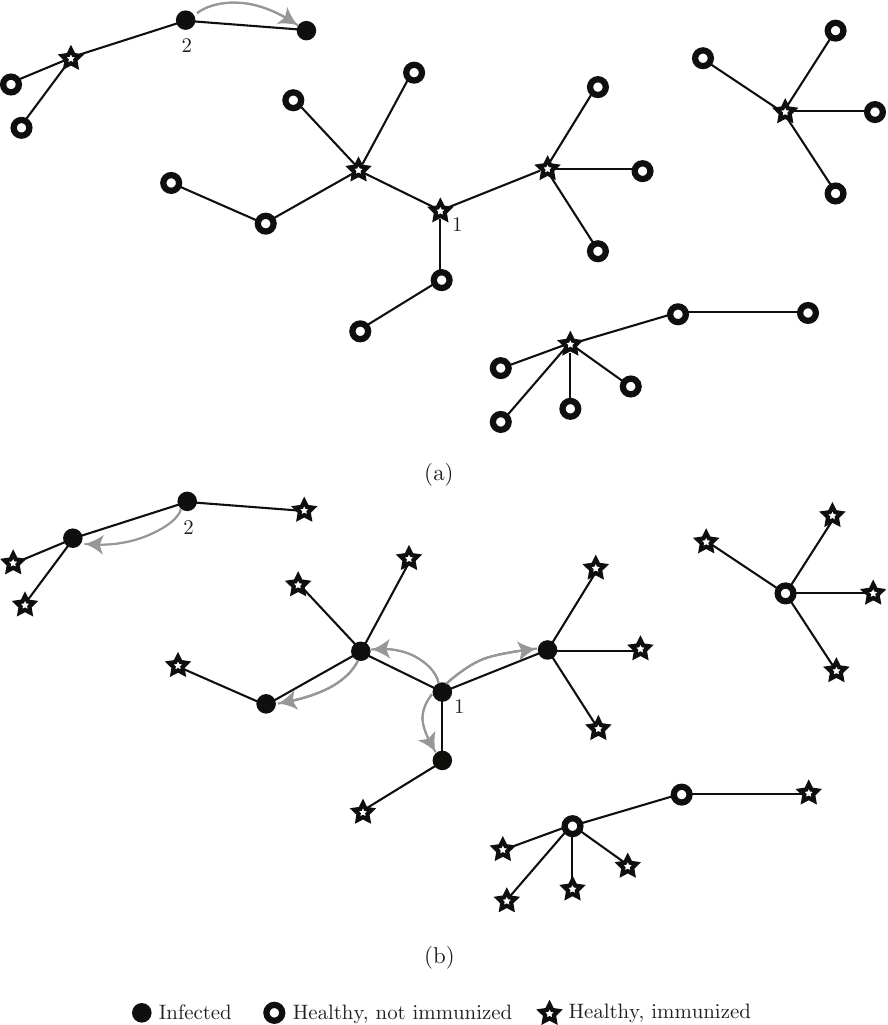}
\caption{Illustration of Theorems \ref{th::::uniqueness} and \ref{th::::thresholdSelfProtection} on a complex network with the cascading process of Fig. \ref{fig:::CascadeExample}: possible equilibrium strategies in (a) a game of total protection and (b) a game of self protection. In (a), we see that the upper-threshold strategy insulates contagion hubs whereas in (b) we see that the lower-threshold strategy insulates periphery nodes and leaves contagion hubs vulnerable.}
  \label{fig:Them4_5}
\end{figure*}

In Fig. \ref{fig:Them4_5}, we illustrate Theorems \ref{th::::uniqueness} and \ref{th::::thresholdSelfProtection} on a complex network. We see how the upper (resp. lower) threshold nature of equilibria in games of total (resp. self) protection affects the spread of cascading failures differently.  

We now state a result comparing the welfare in games of total and self protection. 
\begin{proposition}[Welfare III]
\label{prop::::welfare_III}
   Let $W(\mu) = \sum_d f(d) U_d(\mu(d),\mu)$ be the utilitarian welfare under strategy $\mu$. Specifically, we denote by $W^{tp}(\cdot)$ the utilitarian welfare in a game of total protection and by $W^{sp}(\cdot)$ the utilitarian welfare in a game of self protection, when all other model parameters are held fixed.
 Then $W^{tp}(\mu^*) > W^{sp}(\bar{\mu}^*)$, where $\mu^*$ is the unique equilibrium in a game of total protection and $\bar{\mu}^*$ be the maximum-investment equilibrium in a game of self protection.
 
 \end{proposition}

The above proposition states that the unique equilibrium in a game of total protection welfare-dominates the higher-investment equilibrium in a game of self protection. This result is mainly due to the fact that the return on investment in a game of total protection is higher than in a game of self protection, since it protects against the total risk of failure (not just the intrinsic risk of failure).


An advantage of our informational setting is that we can relate equilibrium behavior to network properties as captured by the edge-perspective degree distribution $\tilde{f}(d)$. We can then ask questions such as ``does a higher level of connectedness\footnote{Note that by a higher level of connectedness, we mean an edge-perspective degree distribution placing higher mass on higher-degree nodes. We do not mean the presence of short paths between any two nodes.} increase or decrease the incentives to invest in protection?" This is examined in the next proposition.

\begin{proposition}[Shifting Degree Distribution]
\label{pr::::compStat_self_acc}
Let $\underline{\mu}^*$ and $\bar{\mu}^*$ be the minimum- and maximum-investment equilibria in a game of self protection, when the edge-perspective degree distribution is $\tilde{f}$. Then, a first-order distributional shift\footnote{Here $\tilde{f}' \succ \tilde{f}$ means that $\tilde{f}'$ first-order stochastically dominates $\tilde{f}$. } $\tilde{f}' \succ \tilde{f}$ results in $\underline{\mu}'^* \preceq \underline{\mu}^* $ and $\bar{\mu}'^* \preceq \bar{\mu}^*$ and thus in $\mc{T}'(\underline{\mu}'^*) \geq \mc{T}(\underline{\mu}^*)$ and $\mc{T}'(\bar{\mu}'^*) \geq \mc{T}(\bar{\mu}^*)$.
\end{proposition}

Thus in a game of self protection, a higher level of connectedness leads to \em lower \em incentives to invest in protection: each of the new maximum- and minimum-investment equilibria are weakly dominated by the corresponding equilibria in the less connected network. The intuition behind this result is that an agent is more likely to be connected to a high-degree neighbor (high contagion risk and unprotected). This increases the agent's cascading failure risk and therefore lowers the incentive to invest in \em self \em protection. We note that in equilibrium, the corresponding neighbor failure probabilities are larger, i.e. $\mc{T}'(\underline{\mu}'^*) \geq \mc{T}(\underline{\mu}^*)$ and $\mc{T}'(\bar{\mu}'^*) \geq \mc{T}(\bar{\mu}^*)$.
  
Note that there is no straightforward analogue to Proposition \ref{pr::::compStat_self_acc} in the case of total protection or networked-risk protection. In fact shifting $\tilde{f}(d)$ may in this case increase the probability of having a protected neighbor or an unprotected one, depending on the extent of the shift in $\tilde{f}(d)$ and on the threshold $d_U$ in the upper-threshold strategy. A shift in $\tilde{f}(d)$ could thus potentially have non-monotone effects.
  
When cascading failures follow a contact process as in the examples of Section \ref{sec:FailProbs_utilities}, it is interesting to study the effect of a change in the infectiousness parameter $r$ on equilibria. The following two propositions illustrate that a change in $r$ has opposite effects, depending on whether the game is one of self protection or total protection.

\begin{proposition}[Varying Infectiousness]
\label{pr::::compStat_r}

Suppose cascading failures follow a contact process with infectiousness parameter $r$, as in the examples of Section \ref{sec:FailProbs_utilities}. Let $\underline{\mu}^*$ and $\bar{\mu}^*$ be the minimum- and maximum-investment equilibria in a game of self protection and let $\mu^*$ be the unique equilibrium in a game of total (or networked-risk) protection. Then, an increase $r'>r$ in infectiousness results in: 
\begin{itemize}
\item{(i)}  $\underline{\mu}'^* \preceq \underline{\mu}^*$ and $\bar{\mu}'^* \preceq \bar{\mu}^*$ and thus in $\mc{T}'(\underline{\mu}'^*) \geq \mc{T}(\underline{\mu}^*)$ and $\mc{T}'(\bar{\mu}'^*) \geq \mc{T}(\bar{\mu}^*)$.

\item{(ii)}  $\mu'^* \succeq \mu^*$ and $r' \mc{T}'(\mu'^*) \geq r \mc{T}(\mu^*)$.
\end{itemize}
\end{proposition}

Part (i) says that in a game of self protection, when cascading failures follow a contact process, a higher level of infectiousness creates \em lower \em incentives for agents to invest in protection: the initial increase in $\mc{T}(\mu)$ caused by higher infectiousness causes an even greater increase in $\mc{T}(\mu)$ as a result of strategic interactions. The situation is very different in a game of total (or networked-risk) protection, as shown in Part (ii), where a higher level of infectiousness creates \em higher \em incentives for agents to invest in protection. This investment in protection is however not enough to counter the increase in $r \mc{T}(\mu)$ caused by a higher level of infectiousness. This is because agents free-ride on the protection provided by others and thus an increase in $r \mc{T}(\mu)$ cannot be completely compensated. 

The next result examines the effect of an increase in the parameter $k$, which governs the extent of the protection resulting from an investment.

\begin{proposition}[Varying the Quality of Protection]
\label{pr::::compStat_self_k}

Let $\underline{\mu}^*$ and $\bar{\mu}^*$ be the minimum- and maximum-investment equilibria in a game of self protection and let $\mu^*$ be the unique equilibrium in a game of total (or networked-risk) protection with parameter $k$. Then, $k' > k$ results in:
\begin{itemize}
\item{(i)} $\underline{\mu}'^* \succeq \underline{\mu}^*$ and $\bar{\mu}'^* \succeq \bar{\mu}^* $, and thus in $\mc{T}'(\underline{\mu}'^*) \leq \mc{T}(\underline{\mu}^*)$ and $\mc{T}'(\bar{\mu}'^*) \leq \mc{T}(\bar{\mu}^*)$.

\item{(ii)}  $\mu'^* \preceq \mu^*$, but $ \mc{T}'(\mu'^*) \leq  \mc{T}(\mu^*)$.
\end{itemize}
\end{proposition}

Thus  in a game of self protection, an increase in the protection quality results in a higher investment and a reduction in a neighbor's probability of failure. Strategic interactions thus further add to the benefits of an improvement in the protection technology. On the contrary, in a game of total (or networked-risk) protection, such an increase in the protection quality results in a lower investment. However, it still results in a reduction of a neighbor's probability of failure, which is due entirely to the increase in protection quality.

\subsection{Endogenizing the Cost of Protection}\label{sec::::global_ext}
So far, we have only examined network effects. That is, a utility function depends on other agents only through the failure probability of one's neighbors. In reality, global feedback effects might also influence an agent's utility. By `global feedback effects', we mean effects that impact an agent's utility in other ways than through its neighbors on the network. For instance, prices of vaccines, computer security solutions or airport security equipment might be affected by demand (i.e. by $\mu$).  Likewise, if protection is provided under the form of insurance\footnote{See, for example, \cite{CyberInsurancePremiumsReuters}: "Cyber insurance premiums rocket after high-profile attacks". Oct 12, 2015. Reuters.  The reader may also see \cite{johnson2011security} and \cite{lelarge2009economic} for some work on insurance provision.}, the insurance premium might depend on the overall failure level in the population, which itself depends on the overall level of investment in protection. Such price feedback effects, in addition to network effects, are also considered in \cite{JacksonZenouHandbook}. \cite{gagnon2015networks} also build a model in which agents' utilities are affected both by their neighbors on a social network and by effects unrelated to that network.

In this section, we introduce such global feedback effects to the model developed in the previous sections. We focus on global feedback through the cost of protection, which can take the form of a price to be paid.

We will introduce the following function, which maps a strategy $\mu$ to the corresponding probability that a randomly-picked agent invests in protection:

\begin{definition}
	\label{def::::Wfunc}
	Let the function $\mc{G}: \mc{M} \rightarrow [0,1]$ be defined as:
	\begin{equation}
		\mc{G}(\mu) = \sum_{d \geq 1}  f(d) \mu(d).
	\end{equation}
\end{definition}
Thus to each strategy $\mu$ corresponds a fraction $\mc{G}(\mu)$ of agents who invest in protection. Furthermore, it is easy to notice that this function $\mc{G}$ increases in $\mu$.

We will explore a setting in which the cost of protection is influenced by global demand. Namely, when the cost of protection depends monotonically on total demand:  $C_{g} = C \cdot g(\mc{G}(\mu))$, where $g(\cdot)$ is either an increasing or a decreasing continuous function of the total fraction of people $ \mc{G}(\mu)$ willing to invest in protection. In the following examples, we outline two situations that can be modeled by the function $g(\cdot)$.\\

\noindent \textbf{Example 1} ($g(\cdot)$ increasing).
This case corresponds to the situation where the product is scarce or there are global congestion effects. For instance, a vaccine might be produced in limited quantity and thus, the more people demand it, the harder it may be to obtain it, which will have an increasing effect on price.\\

\noindent \textbf{Example 2} ($g(\cdot)$ decreasing).
This corresponds to the case of economies of scale. For instance,  a new airport security technology might require significant initial R \& D investments. Producing it in large numbers may thus lead to a lower cost per unit, which may lower the price.\\ 

We will slightly modify a degree-$d$ agent's expected utility function in order to introduce the global feedback effect:
\begin{equation}
	U_d(a,\mu) = -V \cdot \mc{B}(p,q_d(\mathcal{T}(\mu)),a)-C \cdot g(\mc{G}(\mu)) \cdot a.
\end{equation}
Note that the cascading failure probability $q_d(\mathcal{T}(\mu))$ does not depend explicitly on the global fraction of agents who invest in protection, as it is solely driven by network effects, i.e. through an agent's neighborhood. It is also important to mention that the introduction of a global externality does not affect the definition of $\mathcal{T}(\mu)$. The latter function was defined to be the failure probability of a randomly-picked neighbor, which does not depend explicitly on the total fraction of agents investing in protection $\mc{G}(\mu)$.

We will now modify the optimality condition in order to ensure that this fraction $\mc{G}(\mu)$ arises in equilibrium. We can redefine the set of optimal responses as follows:
\begin{definition}
	\label{def::::Sglob}
Let $\mc{S}_d(\mathcal{T}(\mu), \mc{G}(\mu)) \subset [0,1]$ denote the set of optimal responses for a degree-$d$ agent given $\mathcal{T}(\mu)$ and $\mc{G}(\mu)$; i.e.:
\begin{align*}
U_d(1,\mu) > U_d(0,\mu) &\implies \mc{S}_d(\mathcal{T}(\mu), \mc{G}(\mu)) = \{1\};\\
U_d(1,\mu) < U_d(0,\mu)&\implies \mc{S}_d(\mathcal{T}(\mu), \mc{G}(\mu)) = \{0\};\\
U_d(1,\mu) = U_d(0,\mu) &\implies \mc{S}_d(\mathcal{T}(\mu), \mc{G}(\mu)) = [0,1].
\end{align*}
Let $\mc{S}(\mathcal{T}(\mu), \mc{G}(\mu)) \subset \mc{M}$ denote the set of optimal strategies given $\mathcal{T}(\mu)$ and $\mc{G}(\mu)$; i.e.,
\[ \mc{S}(\mathcal{T}(\mu), \mc{G}(\mu)) = \prod_{d \geq 1} \mc{S}_d(\mathcal{T}(\mu), \mc{G}(\mu)). \]
\end{definition}

We now only need to slightly modify the equilibrium condition: 
\begin{definition}[Mean-Field Equilibrium with Endogenized Cost of Protection]
A strategy $\mu^*$ constitutes a mean-field equilibrium (MFE) if $\mu^* \in \mathcal{S}(\mathcal{T}(\mu^*),\mc{G}(\mu^*))$.
\end{definition}

It turns out that the main results that were stated in the previous sections of the paper are robust to the introduction of this global externality. We summarize those more general results in the following proposition.
\begin{proposition}[Network Security Game with Endogenized Cost of Protection]
\label{th::::existence_global}
\quad
\begin{itemize}
\item{(i)}	(Existence): There exists a mean-field equilibrium in the game with endogenized cost or protection. 
\item{(ii)} (Threshold Strategies): The threshold characterization of equilibria is robust to the endogenization of the cost of protection. The equilibrium is of: (1) an upper-threshold nature for a game of total protection and networked-risk protection; (2) a lower-threshold nature for a game of self protection.
\item{(iii)} (Uniqueness): In a game of total protection or of networked-risk protection, the mean-field equilibrium $\mu^*$ is unique if $g(\cdot)$ is an increasing function.
	\end{itemize}
\end{proposition}

As before, there can be multiple equilibria for games of self protection.

\section{Conclusion}
\label{sec::::conclusion}

In this paper, we developed a framework to study the strategic investment in protection against cascading failures in networked systems. Agents connected through a network can fail either intrinsically or as a result of a cascade of failures that may cause their neighbors to fail.  We studied three broad classes of games covering a wide range of applications. We showed that equilibrium strategies are monotone in degree (i.e. in the number of neighbors an agent has on the network) and that this monotonicity is reversed depending on whether (i) an investment in protection insulates an agent against the risk of failure of his neighbors (games of total protection and games of networked-risk protection) or (ii) only against his own intrinsic risk of failure (games of self protection). The first case covers the important examples of vaccination, anti-virus software as well as protection against sexually-transmitted diseases. Here it is the \em more \em connected agents who have higher incentives to invest in protection. The second case, on the other hand,  covers examples such as airport/EU security as well as other types of computer security solutions such as two-factor authentication (2FA). Here it is the \em less \em connected agents who have higher incentives to invest in protection. Our analysis reveals that it is the nature of strategic interactions (strategic substitutes/complements), combined with a network structure that leads to such strikingly different equilibrium behavior in each case, with important implications for the system's resilience to cascading failures. 

Our model is simple and the incomplete information framework that we use allows for a tractable treatment. The unobservability of the network is a credible assumption for many applications. In the applications of vaccination or computer security, agents typically do not know the topology of the social network or email network, for example. They merely know the number of connections that they have. Our equilibrium concept then predicts the behavior of the agents based on their level of interaction with the population (their degree). 

The property that neighbors fail independently imposes a realistic cognitive burden on agents and allows for a tractable way to express an agent's expected cascading failure probability. This property is similar to the local tree-like assumption used in other models such as \cite{lelarge2008local} and is valid for large or relatively sparse networks. In spite of its advantages, this property may no longer be realistic for small or dense networks. However, as long as an agent's cascading failure probability is monotone in degree (which may still be approximately the case, even in some smaller/denser networks), our monotonicity results could hold, at least approximately.

In the case of airport security or EU security, the topology of the network of airports or countries can credibly be known and influence the decisions of the agents. It would be interesting to extend our analysis to the case where agents know the topology of the network. While it is likely that equilibrium behavior would still be monotone in the level of interaction of the agents with the rest of the population, degree centrality may no longer be the appropriate measure. It would be interesting to see if in this case, one could somehow relate equilibrium behavior to some other measure of network centrality as in \cite{acemoglu2013network}. Such extensions are left for future work.

\appendix

\section{Proofs}






\begin{proof}[Proposition \ref{th::::existence}]
Note that we endow $[0,1]$ with the Euclidean topology.

For any $\alpha \in [0,1]$ define the correspondence $\Phi$ by $\Phi(\alpha) = \mc{T}(\mc{S}(\alpha))$.  Any fixed point $\alpha^*$ of $\Phi$, with the corresponding $\mu^* \in \mc{S}(\alpha^*)$ such that $\mc{T}(\mu^*) = \alpha^*$ constitute a MFE. We thus need to show that the correspondence $\Phi$ has a fixed point. We employ Kakutani's fixed point theorem on the composite map $\Phi(\alpha) = \mc{T}(\mc{S}(\alpha))$. 

Kakutani's fixed point theorem requires that $\Phi$ have a compact domain, which is trivial since $[0,1]$ is compact.  Further, $\Phi(\alpha)$ must be nonempty; again, this is straightforward, since both $\mc{S}$ and $\mc{T}$ have nonempty image.  

Next, we show that $\Phi(\alpha)$ has a closed graph.  We first show that $\mc{S}$ has a closed graph, when we endow the set of strategies with the product topology on $[0,1]^\infty$.  This follows easily: if $\alpha_n \to \alpha$, and $\mu_n \to \mu$, where $\mu_n \in \mc{S}(\alpha_n)$ for all $n$, then $\mu_n(d) \to \mu(d)$ for all $d$. Expressing utility as a function of $\alpha$, i.e. $U_d(a,\alpha) = -V \cdot \mc{B}(p,q_d(\alpha),a)-C  \cdot a$, we see that $U_d(1,\alpha)$ and $U_d(0,\alpha)$ are continuous, and it follows that $\mu(d) \in \mc{S}_d(\alpha)$, so $\mc{S}$ has a closed graph.  Note also that with the product topology on the space of strategies, $\mc{T}$ is continuous: if $\mu_n \to \mu$, then $\mc{T}(\mu_n) \to \mc{T}(\mu)$ by the bounded convergence theorem.

To complete the proof that $\Phi$ has a closed graph, suppose that $\alpha_n \to \alpha$, and that $\alpha'_n \to \alpha'$, where $\alpha'_n \in \Phi(\alpha_n)$ for all $n$.  Choose $\mu_n \in \mc{S}(\alpha_n)$ such that $\mc{T}(\mu_n) = \alpha_n'$ for all $n$.  By Tychonoff's theorem, $[0,1]^\infty$ is compact in the product topology; so taking subsequences if necessary, we can assume that $\mu_n$ converges to a limit $\mu$.  Since $\mc{S}$ has a closed graph, we know $\mu \in \mc{S}(\alpha)$.  Finally, since $\mc{T}$ is continuous, we know that $\mc{T}(\mu) = \alpha'$.  Thus $\alpha' \in \Phi(\alpha)$, as required.

Finally, we show that the image of $\Phi$ is convex. Let $\alpha_1, \alpha_2 \in \Phi(\alpha)$ , and choose $\mu_1, \mu_2 \in \mc{S}(\alpha)$ such that $\alpha_1 = \mc{T}(\mu_1)$ and $\alpha_2 = \mc{T}(\mu_2)$. Since $\mc{F}$ is continuous in $\mu$ and since $\mc{T}$ is unique (this follows from Property \ref{ass::::T_unique}), then $\mc{T}$ is continuous in $\mu$. Now since $\mc{S}(\alpha)$ is convex, it follows that for any $\delta \in (0,1)$, 
\begin{eqnarray*}
\delta \mc{T}(\mu_1) + (1 - \delta)\mc{T}(\mu_2) &\in& [\underset{\mu \in \mc{S}(\alpha)}{\operatorname{min}} \mc{T}(\mu), \underset{\mu \in \mc{S}(\alpha)}{\operatorname{max}} \mc{T}(\mu) ] \\
 &=& \Phi(\alpha)  
\end{eqnarray*}
and thus $\delta \alpha_1 + (1- \delta)\alpha_2 \in \Phi(\alpha)$---as required.

By Kakutani's fixed point theorem, $\Phi$ possesses a fixed point $\alpha^*$.  Letting $\mu^* \in \mc{S}(\alpha^*)$ be such that $\mc{T}(\mu^*) = \alpha^*$, we conclude that $\mu^*$ is an MFE.  
\end{proof}

\begin{proof}[Theorem \ref{th::::uniqueness}]

For convenience, since the expected utility $U_d(a,\mu)$ depends on $\mu$ only through $\alpha = \mc{T}(\mu)$, we may write it as a function of $\alpha$ as follows. 

\begin{equation}
	U_d(a,\alpha) = -V \cdot \mc{B}(p,q_d(\alpha),a)-C  \cdot a
\end{equation}

Consider the incremental expected utility for an agent of degree $d$, i.e. 
\begin{eqnarray}
\label{DeltaU}
\Delta  U_d(\alpha)  &=& U_d(1,\alpha) - U_d(0,\alpha)  \\
   &=& -V \cdot (p + (1-p) q_d(\alpha))(1-k) -C - (- V \cdot (p + (1-p)q_d(\alpha)) \nonumber \\ 
 &=& V \cdot \Big( p + (1-p)q_d(\alpha) \Big)k - C \nonumber
\end{eqnarray}

We will first show that any equilibrium is an upper-threshold strategy.

Consider $\Delta  U_d(\alpha)$ as a function of the continuous variable $d$ over the connected support $[1,\infty)$. From (\ref{DeltaU}), we can write
\begin{eqnarray}
\Delta  U_d(\alpha)  &=&  V \cdot \Big( p + (1-p)q_d(\alpha) \Big)k - C \nonumber
\end{eqnarray}

Since $q_d(\alpha)$ is non-decreasing in $d$, for any $\alpha \in (0,1)$, $\Delta U_d(\alpha) $ is a non-decreasing function of $d$. It follows that the inverse image of $(-\infty,0)$ is $\emptyset$ if $\Delta U_1(\alpha)>0$ or an interval $[1,x)$ where $x \geq 1$ otherwise. The integers in such intervals (i.e. $\emptyset \bigcap \mathbbm{N}^+$ or $[1,x) \bigcap \mathbbm{N}^+$) represent the degrees of agents for whom not investing in protection is a strict best response, i.e. $\{d: \mc{S}_d (\alpha) = \{0\} \}$. It follows that the degrees of agents for whom investing in protection is a strict best response (i.e. $\{d: \mc{S}_d (\alpha) = \{1\} \}$) are located at the rightmost extremity of the degree support.

Thus we may write $\mu(d) = 1$, for all $d > d_U$ and $\mu(d) = 0$, for all $d < d_U$. This is valid for any best-responding strategy $\mu$ and it is therefore valid for any equilibrium strategy $\mu^*$.

We now prove equilibrium uniqueness. We prove it in a sequence of steps:

{\em Step 1: For all $d \geq 1$, $\Delta  U_d(\alpha) $ is strictly increasing in $\alpha \in [0,1]$.} This follows directly from Definition \ref{def:qd}.

{\em Step 2: For all $d \geq 1$, and $\alpha' > \alpha$, $\mc{S}_d(\alpha') \succeq \mc{S}_d(\alpha)$.}\footnote{Here the set relation $A\preceq B$ means that for all $x \in A$ and $y \in B$, $x \leq y$.}  This follows immediately from Step 1 and the definition of $\mc{S}_d$ in Definition \ref{def::::S}.

{\em Step 3: If $\mu'$, $\mu$ are strategies such that $\mu'(d) \geq \mu(d)$, then $\mc{T}(\mu') \leq \mc{T}(\mu)$.}  This follows from the fact that $\mc{F}(\mu,\alpha)$ (cf. (\ref{eq::::FtotalProt}) in Definition \ref{def::::totalProt}) is non-increasing in $\mu$ and that it is also continuous in both $\mu$ and $\alpha$. Thus the unique fixed point $\bar{\alpha} = \mc{F}(\mu,\bar{\alpha})$ is non-increasing in $\mu$. Therefore, $\mc{T}(\mu') \leq \mc{T}(\mu)$.

{\em Step 4: Completing the proof.}  So now suppose that there are two mean-field equilibria $\mu^*$ and $\mu'^{*}$, with $\mc{T}(\mu'^{*})=\alpha'^{*} > \alpha^* = \mc{T}(\mu^*)$.  By Step 2, since $\mu^* \in \mc{S}(\alpha^*)$ and $\mu'^{*} \in \mc{S}(\alpha'^{*})$, we have $\mu'^*(d) \geq \mu^{*}(d)$.  By Step 3, we have $\alpha^* =\mc{T}(\mu^*) \geq \mc{T}(\mu'^{*}) = \alpha'^{*}$, a contradiction.  Thus the $\alpha^* = \mc{T}(\mu^*)$ in any MFE must be unique, as required. 

It then follows from the threshold nature of the equilibrium strategy $\mu^*$ that to $\alpha^*$, there corresponds a unique $\mu^* \in \mc{S}(\alpha^*)$ such that $\alpha^* = \mc{T}(\mu^*)$.



\end{proof}

\begin{proof}[Theorem \ref{th::::thresholdSelfProtection}]
For convenience, we do as in the proof of Theorem \ref{th::::uniqueness} and write the expected utility $U_d(a,\alpha)$ as a function of $\alpha$.

In a game of self protection, consider now $\Delta  U_d(\alpha)$ as a function of the continuous variable $d$ over the connected support $[1,\infty)$. From (\ref{eq::::BselfProt}) and (\ref{eq::::U}), we can write
\begin{eqnarray}
\Delta  U_d(\alpha)  &=& U_d(1,\alpha) - U_d(0,\alpha)  \nonumber \\
&=& - V \cdot (p(1-k) + (1-p(1-k)) q_d(\alpha))  -C +V \cdot (p+ (1-p)q_d(\alpha)) \nonumber \\
&=&   V \cdot (pk -pkq_d(\alpha)) - C \nonumber
\end{eqnarray}

Since $q_d(\alpha)$ is non-decreasing in $d$, for any $\alpha \in (0,1)$, $\Delta U_d(\alpha) $ is a non-increasing function of $d$. It follows that the inverse image of $(-\infty,0)$ is an interval $[1,\infty)$ if $\Delta U_1(\alpha)<0$ or an interval $(x,\infty)$ where $x \geq 1$ otherwise. The integers in such intervals (i.e. $[1,\infty) \bigcap \mathbbm{N}^+$ or $(x,\infty) \bigcap \mathbbm{N}^+$) represent the degrees of agents for whom not investing in protection is a strict best response, i.e. $\{d: \mc{S}_d (\alpha) = \{0\} \}$. It follows that the degrees of agents for whom investing in protection is a strict best response (i.e. $\{d: \mc{S}_d (\alpha) = \{1\} \}$) are located at the leftmost extremity of the degree support.

Thus we may write $\mu(d) = 1$, for all $d < d_L$ and $\mu(d) = 0$, for all $d > d_L$. This is valid for any best-responding strategy $\mu$ and it is therefore valid for the equilibrium strategy $\mu^*$.

%
%
\end{proof}



\begin{proof}[Corollary \ref{cor:networkedrisk}]
The result follows from comparing the incremental expected utility of an agent of degree $d$ in the case of networked-risk protection (see (\ref{eq:cor_net_risk}) below) with the one in the case of total protection (see (\ref{DeltaU}) in the proof of Theorem \ref{th::::uniqueness}):
\begin{equation}
\label{eq:cor_net_risk}
\Delta U_d(\alpha) = V(1-p) q_d(\alpha) k - C
\end{equation}
By comparing (\ref{eq:cor_net_risk}) to (\ref{DeltaU}), it is easy to see that the incremental utility of investing in protection is $Vpk>0$ higher in the case of total protection. Otherwise, $\Delta U_d(\alpha)$ is increasing in $d$ and a similar argument as in the proof of Theorem \ref{th::::uniqueness} leads to the conclusion that the equilibrium strategy $\mu^*$ is of an upper-threshold nature. 
\end{proof}

\begin{proof}[Proposition \ref{prop:risk_welfare_I}]

Part (i):

Note that $q_{d}(\mc{T}(\mu^*))$ is non-decreasing in $d$. Thus for $d' > d$, $a_{d'} \in \mu^*(d')$ and $a_d \in \mu^*(d)$, we have
\begin{equation}
\label{eq:::proofWelfare}
U_d(a_d,\mu^*) \geq U_d(a_{d'},\mu^*) \geq U_{d'}(a_{d'},\mu^*)
\end{equation}
where the first inequality follows from $a_d \in \mu^*(d)$, while the second inequality follows from $q_{d}(\mc{T}(\mu^*))$ being non-decreasing in $d$. Thus $U_d(a_d,\mc{T}(\mu^*))$ is non-increasing in $d$.

Part (ii):

From Theorem \ref{th::::thresholdSelfProtection}, the equilibrium strategy $\mu^*$ is non-increasing in $d$. From (\ref{eq::::BselfProt}), it thus follows that for $a_d \in \mu^*(d)$, $\mc{B}(p,q_d(\mc{T}(\mu^*)),a_d)$ is non-decreasing in $d$ (since $\mc{B}$ is non-decreasing in $q_d(\mc{T}(\mu^*))$ and non-increasing in $a_d$).

\end{proof}

\begin{proof}[Proposition \ref{prop::::welfare_II}]

For convenience, we do as in the proof of Theorem \ref{th::::uniqueness} and write the expected utility $U_d(a,\alpha)$ as a function of $\alpha$.

Let $\alpha_l^* = \mc{T}(\mu_l^*)$ and $\alpha_k^* = \mc{T}(\mu_k^*)$.  We then have $\mu_l^* \in \mc{S}(\alpha_l^*)$ and $\mu_k^* \in \mc{S}(\alpha_k^*)$. Then for any $d$, 
\begin{equation}
\label{eq::::proofWelfare}
U_d(a_l,\alpha_l^*) \geq U_d(a_k,\alpha_l^*) \geq U_d(a_k,\alpha_k^*)
\end{equation}
where $a_l \in \mu_l^*(d)$ and $a_k \in \mu_k^*(d)$. 

The first inequality follows from $a_l$ being a best response to $\alpha_l^*$ (i.e. $a_l \in \mu_l^*(d)$) for an agent of degree $d$. The second inequality follows from $U_d$ being decreasing in $\alpha^*$.

Since (\ref{eq::::proofWelfare}) holds for any $d$, all agents have expected utility that is weakly greater in the higher-investment equilibrium $\mu_l^*$. We therefore conclude that $\mu_l^*$ weakly Pareto-dominates $\mu_k^*$.  

\end{proof}

\begin{proof}[Proposition \ref{prop::::welfare_III}]

First note that the expected utilities in all possible cases are
\begin{align*}
U_d^{tot}(1,\mu^*) &= -V \big( p + (1-p) q_d(\mc{T}(\mu^*))\big)(1-k)-C \\
U_d^{tot}(0,\mu^*) &= -V\big( p + (1-p) q_d(\mc{T}(\mu^*))\big) \\
U_d^{s.p.}(1,\bar{\mu}^*) &= -V \big( p(1-k) + (1-p(1-k)) q_d(\mc{T}(\bar{\mu}^*))\big) -C \\
U_d^{s.p.}(0,\bar{\mu}^*) &= -V\big( p + (1-p) q_d(\mc{T}(\bar{\mu}^*)) \big).
\end{align*}

Also note that the incremental utilities from investing in each class of games are
\begin{eqnarray}
\label{eq:DU_tot_welfare3}
\Delta U_d^{tot}(\mu^*) &=& U_d^{tot}(1,\mu^*) - U_d^{tot}(0,\mu^*) \nonumber \\
&=& V \big( p + (1-p)q_d(\mc{T}(\mu^*)) \big)k - C
\end{eqnarray}
and that
\begin{eqnarray}
\label{eq:DU_sp_welfare3}
\Delta U_d^{s.p.}(\bar{\mu}^*) &=& U_d^{tot}(1,\bar{\mu}^*) - U_d^{tot}(0,\bar{\mu}^*) \nonumber \\
&=& V \big( p  -pq_d(\mc{T}(\bar{\mu}^*)) \big)k - C.
\end{eqnarray}

Suppose $Vpk>C$. Then, from (\ref{eq:DU_tot_welfare3}) we see that in a game of total protection $\Delta U_d^{tot}(\mu^*)>0$ for all $d$ and thus $\mu^*(d)=1$ for all $d$. Thus
\begin{equation}
 W^{tot}(\mu^*) = \sum_{d} f(d) U_d^{tot}(1,\mu^*).
 \end{equation}
Also, from (\ref{eq:DU_sp_welfare3}) we see that in a game of self protection not all agents may find it optimal to invest and thus $\bar{\mu}^*$ is some lower-threshold strategy. Therefore,
\begin{eqnarray}
W^{s.p.}(\bar{\mu}^*) &=& \sum_{d < d_L} f(d)  U_d^{s.p.}(1,\bar{\mu}^*)  
+ \sum_{d>d_L} f(d)  U_d^{s.p.}(0,\bar{\mu}^*) \nonumber \\
&& + f(d_L)  \mu(d_L) U_d^{s.p.}(1,\bar{\mu}^*) 
 + f(d_L) (1-\mu(d_L) ) U_d^{s.p.}(0,\bar{\mu}^*). 
\end{eqnarray}

Since $\mu^*  \succeq \bar{\mu}^*$, then $\mc{T}(\mu^*) \leq \mc{T}(\bar{\mu}^*)$. Examining the expressions for the utilities allows us to conclude that $U_d^{s.p.}(1,\bar{\mu}^*) < U_d^{tot}(1,\mu^*)$ and that $U_d^{s.p.}(0,\bar{\mu}^*) < U_d^{tot}(0,\mu^*)<U_d^{tot}(1,\mu^*)$, where the last inequality follows from the fact that $\Delta U_d^{tot}(\mu^*)>0$. Comparing the expressions for the welfare then allows us to conclude that $W^{s.p.}(\bar{\mu}^*) < W^{tot}(\mu^*) $.

Now suppose $Vpk \leq C$. Then in a game of total protection, examining (\ref{eq:DU_tot_welfare3}) tells us that not all agents may find it optimal to invest. $\mu^*$ is thus some upper-threshold  strategy and thus 
\begin{eqnarray}
 W^{tot}(\mu^*) &=& \sum_{d > d_U} f(d) U_d^{tot}(1,\mu^*) 
 + \sum_{d<d_U} f(d) U_d^{tot}(0,\mu^*) \nonumber \\		
&& + f(d_U)  \mu(d_U) U_d^{tot}(1,\mu^*) 
 + f(d_U) (1-\mu(d_U) ) U_d^{tot}(0,\mu^*). 
\end{eqnarray}
Also, in a game of self protection, $\bar{\mu}^*(d)=0$ for all $d$. Indeed $\Delta U^{s.p.}_d(\bar{\mu}^*) < 0$ (see (\ref{eq:DU_sp_welfare3})).  Thus
\begin{eqnarray}
W^{s.p.}(\bar{\mu}^*) &=& \sum_{d} f(d) U_d^{s.p.}(0,\bar{\mu}^*)  	
\end{eqnarray}

Since $\mu^*  \succeq \bar{\mu}^*$, then $\mc{T}(\mu^*) \leq \mc{T}(\bar{\mu}^*)$.
Noting that $U_d^{s.p.}(0,\bar{\mu}^*) < U_d^{tot}(0,\mu^*)$ and that $U_d^{s.p.}(0,\bar{\mu}^*) < U_d^{tot}(0,\mu^*) \leq U_d^{tot}(1,\mu^*)$ for all $d$ such that $\mu^*(d)>0$, we conclude by comparing the expressions for the welfare that $W^{s.p.}(\bar{\mu}^*) < W^{tot}(\mu^*) $. 

Thus for all parameter ranges, $W^{s.p.}(\bar{\mu}^*) < W^{tot}(\mu^*) $. This completes the proof.

\end{proof}

\begin{proof}[Proposition \ref{pr::::compStat_self_acc}]
Let $\mc{F'}(\mu,\alpha)$ and $\mc{F}(\mu,\alpha)$ denote (\ref{eq::::FselfProt}) under $\tilde{f'}$ and $\tilde{f}$ respectively. $q_d(\alpha)$ is non-decreasing in $d$ and we know from Theorem \ref{th::::thresholdSelfProtection} that in a game of self-protection, any equilibrium strategy is a lower-threshold strategy. We therefore only need to consider such strategies. It then follows from (\ref{eq::::FselfProt}) that given any lower-threshold strategy $\mu$, $\mc{F'}(\mu,\alpha) \geq \mc{F}(\mu,\alpha)$ for all $\alpha \in [0,1]$. Since by Property \ref{ass::::T_unique}, (\ref{eq::::FselfProt}) has a single fixed point in $\alpha$ and we conclude that $\mc{T'}(\mu) \geq \mc{T}(\mu)$, where $\mc{T'}(\mu)$ and $\mc{T}(\mu)$ denote (\ref{eq::::Tgeneral}) under $\tilde{f}'$ and $\tilde{f}$ respectively. 

It then follows that 
\begin{eqnarray*}
\Phi'(\alpha) &=& \mc{T'}(\mc{S}(\alpha)) \\
					&\succeq& \mc{T}(\mc{S}(\alpha)) \\
					&=& \Phi(\alpha) 
\end{eqnarray*}

It therefore follows that $\underline{\alpha}'^* = min \{\alpha : \alpha = \Phi'(\alpha)\} \geq min \{ \alpha : \alpha = \Phi(\alpha) \} = \underline{\alpha}^*$ and that $\bar{\alpha}'^* = max \{\alpha : \alpha = \Phi'(\alpha)\} \geq max \{ \alpha : \alpha = \Phi(\alpha) \} = \bar{\alpha}^*$.

Thus, the equilibrium strategies are such that $\underline{\mu}'^* = \mc{S}(\bar{\alpha}'^*) \preceq \mc{S}(\bar{\alpha}^*)  = \underline{\mu}^*$ and $\bar{\mu}'^* = \mc{S}(\underline{\alpha}'^*) \preceq \mc{S}(\underline{\alpha}^*)  = \bar{\mu}^*$. Likewise, $\mc{T}'(\bar{\mu}'^*) = \underline{\alpha}'^* \geq  \underline{\alpha}^* = \mc{T}(\bar{\mu}^*)$ and $\mc{T}'(\underline{\mu}'^*) = \bar{\alpha}'^* \geq \bar{\alpha}^*=\mc{T}(\underline{\mu}^*)$.
 
\end{proof}

\begin{proof}[Proposition \ref{pr::::compStat_r}]

Part(i):

Let $\mc{F'}(\mu,\alpha)$ and $\mc{F}(\mu,\alpha)$ denote (\ref{eq::::FselfProt}) under $r'$ and $r$ respectively. In the case of the contact process described in the examples of Section \ref{sec:FailProbs_utilities}, $q'_d(\alpha) > q_d(\alpha)$ for all $\alpha \in [0,1]$, $d>0$. It then follows from (\ref{eq::::FselfProt}) that given any strategy $\mu$, $\mc{F'}(\mu,\alpha) \geq \mc{F}(\mu,\alpha)$ for all $\alpha \in [0,1]$. Since by Property \ref{ass::::T_unique}, (\ref{eq::::FselfProt}) has a single fixed point, we conclude that $\mc{T'}(\mu) \geq \mc{T}(\mu)$, where $\mc{T'}(\mu)$ and $\mc{T}(\mu)$ denote the correspondence (\ref{eq::::Tgeneral}) under $r'$ and $r$ respectively.

It then follows that 
\begin{eqnarray*}
\Phi'(\alpha) &=& \mc{T'}(\mc{S}(\alpha)) \\
					&\succeq& \mc{T}(\mc{S}(\alpha)) \\
					&=& \Phi(\alpha) 
\end{eqnarray*}

It therefore follows that $\underline{\alpha}'^* = min \{\alpha : \alpha = \Phi'(\alpha)\} \geq min \{ \alpha : \alpha = \Phi(\alpha) \} = \underline{\alpha}^*$ and that $\bar{\alpha}'^* = max \{\alpha : \alpha = \Phi'(\alpha)\} \geq max \{ \alpha : \alpha = \Phi(\alpha) \} = \bar{\alpha}^*$.


Thus, the equilibrium strategies are such that $\underline{\mu}'^* = \mc{S}(\bar{\alpha}'^*) \preceq \mc{S}(\bar{\alpha}^*)  = \underline{\mu}^*$ and $\bar{\mu}'^* = \mc{S}(\underline{\alpha}'^*) \preceq \mc{S}(\underline{\alpha}^*)  = \bar{\mu}^*$. Likewise, $\mc{T}'(\bar{\mu}'^*) = \underline{\alpha}'^* \geq  \underline{\alpha}^* = \mc{T}(\bar{\mu}^*)$ and $\mc{T}'(\underline{\mu}'^*) = \bar{\alpha}'^* \geq \bar{\alpha}^*=\mc{T}(\underline{\mu}^*)$. 



Part (ii):

We prove by contradiction. Suppose $r' \mc{T}'(\mu'^*) < r \mc{T}(\mu^*)$. Then $\mc{S} (\mc{T}'(\mu'^*)) \preceq \mc{S} (\mc{T}(\mu^*))$ and thus $\mu'^* \preceq \mu^*$. Since $\mc{F}'(\mu,\alpha) \geq \mc{F}(\mu,\alpha)$ for any $\mu \in \mc{M}$ and $\alpha \in [0,1]$ and since $\mc{F}'$ and $\mc{F}$ are decreasing in $\mu$, we have that $\mc{F}'(\mu'^*,\alpha) \geq \mc{F}(\mu^*,\alpha)$ for any $\alpha \in [0,1]$. Therefore, $\mc{T}'(\mu'^*) \geq \mc{T}(\mu^*)$
and thus, since $r' > r$, we have that $r' \mc{T}'(\mu'^*) > r \mc{T}(\mu^*)$, a contradiction. We conclude that $r' \mc{T}'(\mu'^*) \geq r \mc{T}(\mu^*)$.

It then follows that $\mc{S} (\mc{T}'(\mu'^*))  \succeq  \mc{S}(\mc{T}(\mu^*))$ and thus $\mu'^* \succeq \mu^*$. 

The result extends to games of networked-risk protection by their structural equivalence to games of total protection (see Corollary \ref{cor:networkedrisk}). This completes the proof.  

\end{proof}

\begin{proof}[Proposition \ref{pr::::compStat_self_k}]

Part (i):

Let $\mc{F'}(\mu,\alpha)$ and $\mc{F}(\mu,\alpha)$ denote (\ref{eq::::FselfProt}) under $k'$ and $k$ respectively. It follows from (\ref{eq::::FselfProt}) that given any strategy $\mu$, $\mc{F'}(\mu,\alpha) \leq \mc{F}(\mu,\alpha)$ for all $\alpha \in [0,1]$. Since under by Property \ref{ass::::T_unique}, (\ref{eq::::FselfProt}) has a single fixed point, we conclude that $\mc{T'}(\mu) \leq \mc{T}(\mu)$, where $\mc{T'}(\mu)$ and $\mc{T}(\mu)$ denote the correspondence (\ref{eq::::Tgeneral}) under $k'$ and $k$ respectively.

It then follows that 
\begin{eqnarray*}
\Phi'(\alpha) &=& \mc{T'}(\mc{S}(\alpha)) \\
					&\preceq& \mc{T}(\mc{S}(\alpha)) \\
					&=& \Phi(\alpha) 
\end{eqnarray*}

It therefore follows that $\underline{\alpha}'^* = min \{\alpha : \alpha = \Phi'(\alpha)\} \leq min \{ \alpha : \alpha = \Phi(\alpha) \} = \underline{\alpha}^*$ and that $\bar{\alpha}'^* = max \{\alpha : \alpha = \Phi'(\alpha)\} \leq max \{ \alpha : \alpha = \Phi(\alpha) \} = \bar{\alpha}^*$.

Thus, $\underline{\mu}'^* = \mc{S}(\bar{\alpha}'^*) \succeq \mc{S}(\bar{\alpha}^*)  = \underline{\mu}^*$ and $\bar{\mu}'^* = \mc{S}(\underline{\alpha}'^*) \succeq \mc{S}(\underline{\alpha}^*)  = \bar{\mu}^*$. Likewise, $\mc{T}'(\bar{\mu}'^*) = \underline{\alpha}'^* \leq  \underline{\alpha}^* = \mc{T}(\bar{\mu}^*)$ and $\mc{T}'(\underline{\mu}'^*) = \bar{\alpha}'^* \leq \bar{\alpha}^*=\mc{T}(\underline{\mu}^*)$. 

Part (ii):

We prove by contradiction. Suppose $ \mc{T}'(\mu'^*) >  \mc{T}(\mu^*)$. Then $\mc{S} (\mc{T}'(\mu'^*)) \succeq \mc{S} (\mc{T}(\mu^*))$ and thus $\mu'^* \succeq \mu^*$. Since $\mc{F}'(\mu,\alpha) \leq \mc{F}(\mu,\alpha)$ for any $\mu \in \mc{M}$ and $\alpha \in [0,1]$ and since $\mc{F}'$ and $\mc{F}$ are decreasing in $\mu$, we have that $\mc{F}'(\mu'^*,\alpha) \leq \mc{F}(\mu^*,\alpha)$ for any $\alpha \in [0,1]$. Therefore, $\mc{T}'(\mu'^*) \leq \mc{T}(\mu^*)$, a contradiction. We conclude that $\mc{T}'(\mu'^*) \leq \mc{T}(\mu^*)$.

It then follows that $\mc{S} (\mc{T}'(\mu'^*))  \preceq  \mc{S}(\mc{T}(\mu^*))$ and thus $\mu'^* \preceq \mu^*$. 

The result extends to games of networked-risk protection by their structural equivalence to games of total protection (see Corollary \ref{cor:networkedrisk}). This completes the proof.  

\end{proof}

\begin{proof}[Proposition \ref{th::::existence_global}] 

Part (i):

The proof is analogous to that of Proposition~\ref{th::::existence}, with only minor modifications.

Denote the function $\mathfrak{T}(\mu) = (\mathcal{T}(\mu),\mc{G}(\mu))$ and let the correspondence $\Psi$ be such that $\Psi(\alpha, \omega) = \mathfrak{T}(\mc{S}(\alpha,\omega))$, with the correspondence $\mc{S}(\alpha, \omega)$ defined as in Definition~\ref{def::::Sglob}.  

First, note that $\Psi$ still has a compact domain $[0,1]\times[0,1]$ and a nonempty image. 

Furthermore, it is also simple to show that $\Psi$ has a closed graph. First, note that $\mc{S}(\alpha, \omega)$ has a closed graph when we endow the set of strategies with the product topology on $[0,1]^{\infty}$. Indeed, choose any $(\alpha_n,\omega_n) \rightarrow (\alpha, \omega)$ and $\mu_n \rightarrow \mu$ such that $\mu_n \in \mc{S}(\alpha_n,\omega_n)$. Then $\mu_n(d) \rightarrow \mu(d)$ for any $d$. Expressing utility as a function of $\alpha$ and $\omega$, i.e. $U_d(a,\alpha,\omega) = -V \cdot \mc{B}(p,q_d(\alpha),a)-C\cdot g(\omega)  \cdot a$,  we note that by the continuity of $U_d(1,\alpha, \omega)$ and $U_d(0,\alpha,\omega)$, it follows that $\mu(d) \in \mc{S}_d(\alpha, \omega)$. Thus, $\mc{S}$ has a closed graph. Note also that with the product topology on the space of strategies, $\mathfrak{T}$ is continuous: by the bounded convergence theorem, both $\mc{T}(\mu_n) \rightarrow\mc{T}(\mu)$ and $\mc{G}(\mu_n) \rightarrow \mc{G}(\mu)$  and therefore it is also true that $\mathfrak{T}(\mu_n) \rightarrow \mathfrak{T}(\mu)$. We now only need to consider the sequences $(\alpha_n,\omega_n) \rightarrow (\alpha, \omega)$ and $(\alpha_n', \omega_n') \rightarrow (\alpha', \omega')$ where $(\alpha_n', \omega_n') \in \Psi(\alpha_n, \omega_n)$. By choosing $\mu_n \in \mc{S}(\alpha_n, \omega_n)$ such that $\mc{T}(\mu_n) = \alpha_n'$ and $\mc{G}(\mu_n) = \omega_n'$, and by the same argument as in the proof of Proposition \ref{th::::existence}, we can conclude that $(\alpha', \omega') \in \Psi(\alpha, \omega)$,  as desired.

Finally, the image of $\Psi$ is convex. Indeed, $\mathfrak{T}(\mu)$ is continuous in $\mu$. Furthermore, $\mc{S}(\alpha, \omega)$ is convex (which follows from convexity of $\mc{S}_d(\alpha, \omega)$ for any $d$). Convexity of the image of $\Psi$ thus follows from an argument analogous to that presented in the proof of Proposition \ref{th::::existence}.

By Kakutani's fixed point theorem, $\Psi$ has a fixed point $(\alpha^*, \omega^*)$. Letting $\mu^* \in \mc{S}(\alpha^*, \omega^*)$ be such that $\mathfrak{T}(\mu^*) = (\alpha^*, \omega^*)$, we conclude that $\mu^*$ is an MFE.



Part (ii):

Note that the incremental expected utilities for an agent of degree $d$ in games of total and self protection are respectively:
\begin{equation}
	\label{eq::::incr_util_global_tp}
	\Delta U_d(\mu) = V \cdot \left(p+(1-p) q_d(\mc{T}(\mu))\right) k - C g(\mc{G}(\mu))
\end{equation} 
and
\begin{equation}
	\label{eq::::incr_util_global_sp}
	\Delta U_d(\mu) =V \cdot (pk - pk q_d(\mc{T}(\mu))) - C g(\mc{G}(\mu))
\end{equation} 
It is obvious that for any given $\mc{T}(\mu)$ and $\mc{G}(\mu)$, these functions preserve the properties (i.e. monotonicity in $d$) that were discussed in the proofs of Theorems~\ref{th::::uniqueness} and ~\ref{th::::thresholdSelfProtection}. The threshold nature of equilibria is thus maintained. 
By an argument analogous to that of the proof of Corollary \ref{cor:networkedrisk}, it also follows that a game of networked-risk protection is structurally equivalent to a game of total protection and thus the upper-threshold nature also follows in that case. 



Part (iii):

For convenience, since the expected utility $U_d(a,\mu)$ depends on $\mu$ only through $\alpha = \mc{T}(\mu)$ and $\omega = \mc{G}(\mu)$, we may write it as a function of $\alpha$ and $\omega$ as follows:

\begin{equation}
	U_d(a,\alpha,\omega) = -V \cdot \mc{B}(p,q_d(\alpha),a)-C\cdot g(\omega)  \cdot a
\end{equation}

For a game of total protection, the incremental expected utility for an agent of degree $d$ is
\begin{equation}
	\label{eq::::incr_util_global}
	\Delta U_d(\alpha, \omega) = V \cdot \left(p+(1-p) q_d(\alpha)\right) k - C \cdot g(\omega)
\end{equation} 
We will consider the case when $g(\omega)$ is an increasing function. 
As in the proof of Theorem~\ref{th::::uniqueness}, we will conduct the analysis in 4 steps.

{\em Step 1:} For all $d \geq 1$, $\Delta U_d(\alpha, \omega)$ is strictly increasing in $\alpha \in [0,1]$ and strictly decreasing in $\omega \in [0,1]$. 

{\em Step 2:} Notice that $\alpha$ and $\omega$ are moving $\Delta U_d(\alpha, \omega)$ in opposite directions. Hence, if both $\alpha$ and $\omega$ increase, we cannot conclude anything about the change in $\mc{S}_d(\alpha, \omega)$. However for $\alpha' > \alpha$ and $\omega' > \omega$ it holds $\mc{S}_d(\alpha', \omega) \succeq \mc{S}_d(\alpha, \omega')$.

{\em Step 3:} For any strategies $\mu', \mu$ such that $\mu'(d) \geq \mu(d), \forall d \geq 1$, then $\mc{G}(\mu')\geq \mc{G}(\mu)$ and $\mc{T}(\mu') \leq\mc{T}(\mu)$. As we have noted before, the global externality does not have a direct impact on $\mc{F}(\mu,\alpha)$ and thus the behavior of $\mc{T}(\mu)$ remains as in step 3 of the proof of Theorem~\ref{th::::uniqueness}.

{\em Step 4:} Suppose that there are two mean-field equilibria $(\mu^*, \alpha^*, \omega^*)$ and $(\mu'^*,\alpha'^*, \omega'^*)$. Without loss of generality assume that $\alpha'^* > \alpha^*$. We need to consider two cases. First, if $\omega'^* \leq \omega^*$, then it is true that $\mc{S}(\alpha'^*,\omega'^*) \succeq \mc{S}(\alpha^*, \omega^*)$. As $\mu^* \in \mc{S}(\alpha^*,\omega^*)$ and $\mu'^* \in \mc{S}(\alpha'^*,\omega'^*)$, then it follows that $\mu'^* \succeq \mu^*$. However that leads to the contradiction: $\alpha^* = \mc{T}(\mu^*) \geq \mc{T}(\mu'^*)=\alpha'^*$. Finally consider the case of $\omega'^* > \omega^*$. By Proposition \ref{th::::existence_global}(ii), due to the threshold nature of the equilibrium, the equilibrium strategies can be ordered as either $\mu'^* \succeq \mu^*$ or $\mu'^* \preceq \mu^*$. If $\mu'^* \preceq \mu^*$ then $\omega^* = \mc{G}(\mu^*) \geq \mc{G}(\mu'^*)=\omega'^*$, which is a contradiction. If $\mu'^* \succeq \mu^*$, it follows that $\alpha^* = \mc{T}(\mu^*) \geq \mc{T}(\mu'^*)=\alpha'^*$ and we arrive at a contradiction.  

Thus, we showed that in a game of total protection with both endogenized cost (with $g(\cdot)$ increasing), any MFE must be unique. Uniqueness in the case of a game of networked-risk protection follows by its structural equivalence to a game of total protection (Corollary \ref{cor:networkedrisk}).
\end{proof}

\newpage
\bibliographystyle{elsarticle-harv}
\bibliography{SysRisk_bibfile}

\begin{thebibliography}{31}
\expandafter\ifx\csname natexlab\endcsname\relax\def\natexlab#1{#1}\fi
\expandafter\ifx\csname url\endcsname\relax
  \def\url#1{\texttt{#1}}\fi
\expandafter\ifx\csname urlprefix\endcsname\relax\def\urlprefix{URL }\fi

\bibitem[{Acemoglu et~al.(2016)Acemoglu, Malekian, and
  Ozdaglar}]{acemoglu2013network}
Acemoglu, D., Malekian, A., Ozdaglar, A., 2016. Network security and contagion.
  Journal of Economic Theory 166, 536--585.

\bibitem[{Acemoglu et~al.(2015)Acemoglu, Ozdaglar, and
  Tahbaz-Salehi}]{AcemogluSystemic}
Acemoglu, D., Ozdaglar, A., Tahbaz-Salehi, A., 2015. Systemic risk and
  stability in financial networks. The American Economic Review 105~(2),
  564--608.

\bibitem[{Arribasa and Urbanoa(2014)}]{arribasa2014local}
Arribasa, I., Urbanoa, A., 2014. Local coordination and global congestion in
  random networks. Tech. rep., University of Valencia, ERI-CES.

\bibitem[{Aspnes et~al.(2005)Aspnes, Chang, and
  Yampolskiy}]{aspnes2005inoculation}
Aspnes, J., Chang, K., Yampolskiy, A., 2005. Inoculation strategies for victims
  of viruses and the sum-of-squares partition problem. In: Proceedings of the
  sixteenth annual ACM-SIAM symposium on Discrete algorithms. Society for
  Industrial and Applied Mathematics, pp. 43--52.

\bibitem[{Balthrop et~al.(2004)Balthrop, Forrest, Newman, and
  Williamson}]{Balthrop}
Balthrop, J., Forrest, S., Newman, M., Williamson, M., 2004. Technological
  networks and the spread of computer viruses. Scientific Reports 304,
  527--529.

\bibitem[{Blume et~al.(2013)Blume, Easley, Kleinberg, Kleinberg, and
  Tardos}]{blume2011network}
Blume, L., Easley, D., Kleinberg, J., Kleinberg, R., Tardos, {\'E}., 2013.
  Network formation in the presence of contagious risk. ACM Transactions on
  Economics and Computation 1~(2), 6.

\bibitem[{Cabrales et~al.(2014)Cabrales, Gottardi, and
  Vega-Redondo}]{cabrales2014risk}
Cabrales, A., Gottardi, P., Vega-Redondo, F., 2014. Risk-sharing and contagion
  in networks. SSRN 2425558.

\bibitem[{Cerdeiro et~al.(2015)Cerdeiro, Dziubi{\'n}ski, and
  Goyal}]{cerdeiro2015contagion}
Cerdeiro, D., Dziubi{\'n}ski, M., Goyal, S., 2015. Contagion risk and network
  design. SSRN 2619022.

\bibitem[{Dziubi{\'n}ski and Goyal(2017)}]{GoyalDziubinski}
Dziubi{\'n}ski, M.~K., Goyal, S., 2017. How do you defend a network?
  Theoretical Economics 12~(1), 331--376.

\bibitem[{Elliott et~al.(2014)Elliott, Golub, and
  Jackson}]{JacksonGolubElliott}
Elliott, M., Golub, B., Jackson, M.~O., 2014. Financial networks and contagion.
  The American economic review 104~(10), 3115--3153.

\bibitem[{Gagnon and Goyal(2017)}]{gagnon2015networks}
Gagnon, J., Goyal, S., 2017. Networks, markets and inequality. American
  Economic Review 107~(1), 1--30.

\bibitem[{Galeotti et~al.(2010)Galeotti, Goyal, Jackson, Vega-Redondo, and
  Yariv}]{Galeotti10}
Galeotti, A., Goyal, S., Jackson, M.~O., Vega-Redondo, F., Yariv, L., 2010.
  Network games. Review of Economic Studies. 77, 218--244.

\bibitem[{Galeotti and Rogers(2013)}]{GaleottiRogers}
Galeotti, A., Rogers, B.~W., 2013. Strategic immunization and group structure.
  American Economic Journal: Microeconomics 5~(2), 1--32.

\bibitem[{Goyal and Vigier(2015)}]{GoyalVigier}
Goyal, S., Vigier, A., 2015. Interaction, protection and epidemics. Journal of
  Public Economics 125, 64--69.

\bibitem[{Heal et~al.(2006)Heal, Kearns, Kleindorfer, and
  Kunreuther}]{IDSHealKunreutherKearns}
Heal, G., Kearns, M., Kleindorfer, P., Kunreuther, H., 2006. Interdependent
  security in interconnected networks.

\bibitem[{Heal and Kunreuther(2004)}]{IDSHealKunreuther}
Heal, G., Kunreuther, H., 2004. Interdependent security: A general model. Tech.
  rep., National Bureau of Economic Research.

\bibitem[{Jackson(2008)}]{JacksonBook}
Jackson, M.~O., 2008. Social and economic networks. Princeton University Press,
  NJ.

\bibitem[{Jackson and Yariv(2007)}]{JacksonYariv}
Jackson, M.~O., Yariv, L., 2007. Diffusion of behavior and equilibrium
  properties in network games. American Economic Review 97~(2), 92--98.

\bibitem[{Jackson and Zenou(2014)}]{JacksonZenouHandbook}
Jackson, M.~O., Zenou, Y., 2014. Games on networks. Handbook of Game Theory 4,
  (Peyton Young and Shmuel Zamir, eds.).

\bibitem[{Johnson et~al.(2011)Johnson, B{\"o}hme, and
  Grossklags}]{johnson2011security}
Johnson, B., B{\"o}hme, R., Grossklags, J., 2011. Security games with market
  insurance. In: Decision and Game Theory for Security. Springer, pp. 117--130.

\bibitem[{Leduc(2014)}]{MLeducThesis}
Leduc, M.~V., 2014. Mean-field models in network game theory. Ph.D. thesis,
  Stanford University.

\bibitem[{Leduc et~al.((forthcoming))Leduc, Jackson, and
  Johari}]{leduc2015pricing}
Leduc, M.~V., Jackson, M.~O., Johari, R., (forthcoming). Pricing and referrals
  in diffusion on networks. Games and Economic Behavior.

\bibitem[{Lelarge and Bolot(2008{\natexlab{a}})}]{lelarge2008local}
Lelarge, M., Bolot, J., 2008{\natexlab{a}}. A local mean field analysis of
  security investments in networks. In: Proceedings of the 3rd international
  workshop on Economics of networked systems. ACM, pp. 25--30.

\bibitem[{Lelarge and Bolot(2008{\natexlab{b}})}]{Bolot}
Lelarge, M., Bolot, J., 2008{\natexlab{b}}. Network externalities and the
  deployment of security features and protocols in the internet. SIGMETRICS'08.

\bibitem[{Lelarge and Bolot(2009)}]{lelarge2009economic}
Lelarge, M., Bolot, J., 2009. Economic incentives to increase security in the
  internet: The case for insurance. In: INFOCOM 2009, IEEE. IEEE, pp.
  1494--1502.

\bibitem[{{Reuters}(12 October 2015)}]{CyberInsurancePremiumsReuters}
{Reuters}, 12 October 2015. Cyber insurance premiums rocket after high-profile
  attacks.

\bibitem[{{Reuters}(27 August 2015)}]{USvaccinationRatesReuters}
{Reuters}, 27 August 2015. U.s. vaccination rates high, but pockets of
  unvaccinated pose risk.

\bibitem[{Rosas-Casals et~al.(2007)Rosas-Casals, Valverde, and
  Sol\'{e}}]{RosasCasals}
Rosas-Casals, M., Valverde, S., Sol\'{e}, R.~V., 2007. Topological
  vulnerability of the european power grid under errors and attacks.
  International Journal of Bifurcation and Chaos 17~(7), 2465--2475.

\bibitem[{{The Economist}(4 February 2015)}]{economist_rand_paul}
{The Economist}, 4 February 2015. Rand paul on vaccination: Resorting to
  freedom.

\bibitem[{{The Economist}(5 February 2015)}]{economist_vaccination}
{The Economist}, 5 February 2015. Politics and vaccinations: What experts say,
  and what people hear.

\bibitem[{Wang et~al.(2010)Wang, Scaglione, and Thomas}]{Wang}
Wang, Z., Scaglione, A., Thomas, R.~J., 2010. The node degree distribution in
  power grid and its topology robustness under random and selective node
  removals. 2010 IEEE International Conference on Communications Workshops
  (ICC), 1--5.

\end{thebibliography}

\end{document}